\documentclass[12pt]{article}
\usepackage[bookmarks=false]{hyperref}
\usepackage{fullpage,doublespace}
\usepackage{bm}
\usepackage{amsmath,amssymb,amsthm}
\usepackage{mathtools}
\usepackage[authoryear]{natbib}
\usepackage{lineno}
\usepackage{times}
\usepackage{booktabs} 
\mathtoolsset{showonlyrefs}
\usepackage{graphicx}
\usepackage{url}
\usepackage{multirow}
\usepackage[labelformat=simple]{subcaption}
\usepackage[toc,title,page]{appendix}
\usepackage{hyperref}
\usepackage{enumitem}
\newlist{DS}{enumerate}{1}
\setlist[DS]{label=\textbf{Design \arabic*.},leftmargin=*}

\DeclareMathOperator{\vect}{vec}
\DeclareMathOperator{\diag}{diag}
\DeclareMathOperator{\tr}{tr}

\newcommand{\bU}{ \mbox{\bf U}}
\newcommand{\X}{\bm{X}}

\newcommand{\lcbk}{\left\{}
\newcommand{\rcbk}{\right\}}
\newcommand{\lsbk}{\left[}
\newcommand{\rsbk}{\right]}
\newcommand{\lpth}{\left(}
\newcommand{\rpth}{\right)}

\newenvironment{amatrix}[1]{%
  \left(\begin{array}{@{}*{#1}{c}|c@{}}
}{%
  \end{array}\right)
}

\newcommand{\bx}{ \mbox{\bf x}}
\newcommand{\bX}{ \mbox{\bf X}}
\newcommand{\bB}{ \mbox{\bf B}}

\newcommand{\bz}{ \mbox{\bf z}}

\newcommand{\bV}{ \mbox{\bf V}}

\newcommand{\bD}{ \mbox{\bf D}}
\newcommand{\bM}{ \mbox{\bf M}}

\newcommand{\bT}{ \mbox{\bf T}}

\newcommand{\iid}{\stackrel{iid}{\sim}}
\newcommand{\indep}{\stackrel{indep}{\sim}}

\newcommand{\calD}{{\cal D}}

\newcommand{\beq}{ \begin{equation}}
\newcommand{\eeq}{ \end{equation}}
\newcommand{\beqn}{ \begin{eqnarray}}
\newcommand{\eeqn}{ \end{eqnarray}}

\begin{document}  %\linenumbers

\begin{center}
{\Large Bayesian Non-parametric Quantile Process Regression and Estimation of Marginal Quantile Effects}\\\vspace{6pt}
{\large Steven G. Xu\footnote{Department of Statistics, North Carolina State University, Raleigh, NC, 27695, USA} and Brian J. Reich$^1$}\\
%\today
\date{}
\end{center}

\begin{abstract}\begin{singlespace}
\noindent Flexible estimation of multiple conditional  quantiles is of interest in numerous applications, such as studying the effect of pregnancy-related factors on low and high birth weight. We propose a Bayesian non-parametric method to simultaneously estimate non-crossing, non-linear quantile curves. We expand the conditional distribution function of the response in I-spline basis functions where the covariate-dependent coefficients are modeled  using neural networks. By leveraging the approximation power of splines and neural networks, our model can approximate any continuous quantile function. Compared to existing models, our model estimates all rather than a finite subset of quantiles, scales well to high dimensions, and accounts for estimation uncertainty. While the model is arbitrarily flexible, interpretable marginal quantile effects are estimated using accumulative local effect plots and variable importance measures. A simulation study shows that our model can better recover quantiles of the response distribution when the data is sparse, and an analysis of birth weight data is presented. \end{singlespace}\end{abstract}

\newpage

\section{Introduction}\label{s:intro}
Quantile regression (QR) models a conditional quantile as a function of covariates. It allows one to analyze the statistical relationship between covariates and non-central parts of the conditional response distribution. However, when QR is fitted separately for inference on multiple levels, the natural ordering among different quantiles cannot be ensured, and the estimated quantiles are subject to cross. Quantile crossing can be alleviated by solving a constrained optimization problem \citep{Bondell2010,LiuWu2011} when only a grid of quantiles is modeled, but estimates based on these methods can be sensitive to the number and location of the chosen quantile grids.

Simultaneous QR (SQR) allows inference on all quantiles by specifying the full quantile process. It encourages strength borrowing across proximate quantile levels through an unified modeling approach. SQR was first proposed by \citet{he1997} who assumes a linear heteroscedastic regression model for the response. Subsequently, linear SQR models that impose fewer restrictions on the quantile function have been developed \citep[e.g.,][]{ReichSmith2013,Yuan2017,YangTokdar2017}. These approaches enjoy great interpretability by allowing rate-of-change interpretation of quantile-dependent coefficients but cannot accommodate quantile curves with complex non-linear trends. Furthermore, they are not suitable for high-dimensional problems since they do not implicitly account for interaction effects.

Non-linear SQR models are a minority in the current quantile regression literature, and existing approaches suffer from apparent shortcomings. \citet{Cannon2018} proposed to model the quantile process using a feed-forward neural network. They pre-specify a set of quantile levels and treat them as a monotone covariate in the model to enforce monotonicity of the quantile function. Although their approach elegantly avoids quantile crossing, additional quantiles outside the pre-specified range have to be estimated via extrapolation. \cite{DasGhoshal2018} model the quantile process as a weighted sum of B-spline basis functions of the quantile level; the weights are further expanded by tensor products of B-spline series expansion of each covariate, and order constraints are imposed on the spline coefficients to ensure non-crossing. Although their method models the full quantile process, it does not scale well to high dimensions since the number of parameters grows exponentially with the number of covariates. Non-linear quantile process can also be estimated by inverting (or integrating then inverting) any valid estimate of the conditional distribution function. Das and Ghoshal (2018) proposed a model on the conditional cumulative distribution function (CDF) of similar form to their aforementioned quantile process model. Consequently, their CDF model also suffers from computational intractability in high dimensions. Furthermore, their model is not constrained properly to estimate a bona fide density, which often leads to poor numerical performance. \citet{IzbickiLee2016} projected the conditional probability density function (PDF) onto data-dependent eigenfunctions of a kernel-based operator. Their model scales well to high dimensions and estimates a bona fide density, but the resulting quantile surfaces are often not smooth. Recently, SQR models that leverage the advancement of deep learning have also been developed \citep[e.g.,][]{kim2021}, but the primary focus of these models is on prediction rather than inference.

In this paper, we propose a novel treatment to non-linear SQR by specifying a Bayesian non-parametric model on the conditional distribution. While there exist other conditional distribution regression models
\citep[e.g.,][]{holmes2012,li2019}, we model the conditional CDF using an I-spline basis expansion; the spline coefficients are modeled as functions of covariates using neural networks with specific output activation functions that ensure the model represent a bona fide CDF. We choose to model the distribution function instead of the quantile process because the former permits analytic derivation of the likelihood function and therefore efficient MCMC sampling of the posterior, and for a non-parametric regression the span of potential models is the same in both cases. A spline-based model ensures the estimated conditional CDF, and therefore the estimated conditional quantile function is smooth, and the neural networks allow incorporation of complex covariate effects on the response distribution. We name this method ``QR Using I-spline Neural Network (QUINN)". QUINN provides several improvements over existing non-linear SQR models \citep{IzbickiLee2016,Cannon2018,DasGhoshal2018}. Instead of treating the quantile levels as a monotone covariate, QUINN specifies the full CDF such that all quantiles, rather than only a subset, can be modeled without extrapolation. By imposing proper constraints on the spline coefficient functions, we ensure that the estimate is a bona fide CDF, leading to more accurate estimation of the quantile process; directly constraining the model also avoids the need of an post hoc normalization method which often renders the final quantile estimates unsmooth. Finally, by expanding the covariate-dependent coefficient functions using FNN rather than tensor products of splines, we greatly reduce the dimensionality of the parameter space so that it only scales linearly, instead of exponentially, in the number of covariates. A relevant but distinct work is that of \citet{smith2015} who proposed a semi-parametric framework based on I-spline basis expansion for a simultaneous estimation of linear quantile planes. In contrast, QUINN models the conditional distribution non-parametrically and allows simultaneous estimation of arbitrary quantile surfaces. 

A disadvantage of modeling the conditional CDF non-parametrically is that covariate effects on different quantiles are not self-explanatory. This is common for black box supervised learning models that sacrifice transparency for flexibility. To overcome this challenge, model-agnostic methods \citep{ribeiro2016} have been developed to extract interpretation from any supervise learning model. A recent contribution is made by \citet{ApleyZhu2020} who proposed accumulated local effect (ALE) plot to visualize main and second-order interaction effects of a black box supervised learning model. Their method produces reliable characterization of the covariate effects on the predicted response in a computationally efficient way. In this paper, we show that ALE plots can be applied to visualize covariate effects on predicted quantiles. We also present ways to estimate feature importance of marginal quantile effects of QUINN. 

The motivating example is a study analyzing the effect of pregnancy related and demographic factors on the distribution of birth outcomes. Low birth weight (LBW) is defined as weight less than 2.5kg. It is a leading cause of prenatal and neonatal deaths, and births of underweight infants result in long-term medical and economic costs. High birth weight (HBW) is defined as weight greater than 4kg. It is also an emerging public health issue worldwide. Overweight infants are subject to increased risk of health problems after birth, such as obesity in early childhood. QR is a natural approach to understand the determinants of LBW and HBW by modeling the lower and upper quantiles of the birth weight distribution. Examples are the separate QR approach by \citet{Abrevaya2001} and SQR approach by \cite{TokdarKadane2012}. These works all assume a linear regression model, which will mischaracterize effects that are non-linear \citep{ngwira2015}. In this paper, we apply QUINN to the 2019 U.S. Natality Data Set \citep{nchs2019} to flexibly model different quantiles of the birth weight distribution, with a primary focus on identifying the influential factors of LBW and HBW.

\section{Methods}\label{s:method}
Denote $\bX=(X_1,...,X_d)\in\mathcal{X}$ as the covariate vector and $Y$ as the scalar response. We are interested in approximating the quantile process of the response given the covariates $Q_Y(\tau|\bX=\bx)$ for quantile level  $\tau\in(0,1)$ and all $\bx\in\mathcal{X}$. If $Q_Y(\tau|\bx)$ is continuous and monotonically increasing in $\tau$, then for any $\bx\in\mathcal{X}$ the conditional CDF is  $F_Y(y|\bx)=Q^{-1}_Y(\tau|\bx)$. Thus, the monotonicity constraint of $Q_Y(\tau|\bx)$ can be naturally accounted for by specifying a valid model on $F_Y(y|\bx)$ and then inverting it. Our method requires the response variable to have a lower and upper bound, which we achieve by introducing a transformed variable $Z = g(Y)$ for some monotonic function $g$ that maps to the unit interval. In this section, we will outline our method for approximating quantile process of the transformed response $Q_Z(\tau,\bx)$ whose estimate can then be back-transformed to an estimate of $Q_Y(\tau|\bx)$.

\subsection{Density regression using shape-constrained splines}
\label{s:method:isp}
We propose to model the conditional density of $Z$ given $\bx$ using shape-constrained regression splines. Specifically, we model the conditional PDF using M-splines, and the conditional CDF using I-splines. The M-spline family with degree $r$ is a set of $r+p+1$ piecewise polynomials of degree $r$ having properties of non-negativity and unit integral. In other words, each basis function has the properties of a PDF. Let  $\bT=\{t_1,t_2,\dots,t_{p+2r}\}$ be an ordered sequence of knots such that $t_1=...=t_r=0$, $t_{p+r+1}=...=t_{p+2r}=1$, and $t_{r+k}-t_{r+k-1}=1/p$, $k\in[1,p]$. Let $\{M_{m,r}(\cdot|\bT)\}_{m=1}^{r+p+1}$ be the set of basis functions. A convex combination of M-spline basis functions, i.e., 
\begin{equation*}
    \sum_{m=1}^{r+p+1}\theta_mM_{m,r}(\cdot|\bT)\ \text{s.t.} \ \theta_m\ge0 \ \forall m\ \text{and} \ \sum_{m=1}^{r+p+1}\theta_m=1,
\end{equation*}
is a valid model for a PDF with support on $[0,1]$. The shape of the modeled PDF can be further controlled by placing additional constraints on the coefficients $\theta_m$. For example, setting $\theta_{r+p+1}=0$ will force the M-spline to return to 0 at unity.

I-splines are defined as the integral of M-splines
\begin{equation*}
    I_{m,r}(x|\bT)=\int_0^x M_{m,r}(u|\bT)du, \ m=1,...,r+p+1
\end{equation*}
and are piecewise polynomials of degree $r+1$. Since M-splines are non-negative and integrate to 1, I-splines are monotonically non-decreasing with range $I_{m,r}(0|\bT)=0$ and $I_{m,r}(1|\bT)=1$ for all $m$. Thus, a convex combination of I-spline basis functions, i.e.,
\begin{equation}\label{e:isp-cdf}
    \sum_{m=1}^{r+p+1}\theta_mI_{m,r}(\cdot|\bT)\ \text{s.t.} \ \theta_m\ge0 \ \forall m\ \text{and} \ \sum_{m=1}^{r+p+1}\theta_m=1,
\end{equation}
is a valid model for a CDF with support on the unit interval.
%\begin{equation*} \sum_{m=1}^{r+p-1}\theta_{m}M_{m,r}(z|\bT), \ \theta_{m}\ge0,\ \sum_{m=1}^{r+p-1}\theta_{m}=1  \label{e:0} \end{equation*}

The use of shape-constrained regression splines offers an attractive solution to density estimation problems. Many theoretical works have shown the approximation power of non-negative splines and monotone splines. For example, \citet{beatson1982} shows that as the number of knots increases, the space of non-negative splines converges to the space of non-negative continuous functions almost as quickly as unconstrained splines. \citet{chui1980} show an analogous result for monotonic splines on approximating continuous monotonic functions. Through numerical studies, \citet{abrahamowicz1992} show that the asymptotic theories are not affected by the addition of simplex constraint, and that M- and I-splines yield satisfactory accuracy in density regression.  %the same parameters
%\begin{equation*}
%\sum_{m=1}^{r+p-1}\theta_{m}I_{m,r}(z|\bT) =  \int_L^z\lsbk \sum_{m=1}^{r+p-1}\theta_{m}M_{m,r}(u|\bT)\rsbkdu.
%\end{equation*}

\subsection{QR using I-splines and neural network (QUINN)}
\label{s:method:quinn}
Let $F_Z(z|\bx)$ denote the conditional CDF of the transformed response variable $Z$ given $\bx$. % and $\{I_{m,r}(\cdot|\bT)\}_{m=1}^{r+p-1}$ be the I-spline basis with degree $r$ and knots vector $\bT$. 
Following \eqref{e:isp-cdf}, a flexible model for $F_Z(z|\bx)$ can be expressed as 
\begin{equation*}\label{e:1}
    F_Z(z|\bx,\mathcal{W})=\sum_{m=1}^{r+p+1}\theta_{m}(\bx,\mathcal{W})I_{m,r}(z|\bT) \ \text{s.t.} \ \theta_m(\bx,\mathcal{W})\ge0 \ \forall m\ \text{and} \ \sum_{m=1}^{r+p+1}\theta_m(\bx)=1
\end{equation*}
where the covariates affect the conditional CDF through the spline coefficient functions $\theta_m(\bx,\mathcal{W})$ parametrized by $\mathcal{W}$. The coefficient functions govern the covariate effect on the conditional CDF and therefore should be flexible enough to capture complex non-linear trends and allow for high-order interaction effects. They also need to be properly constrained so that $F_Z(z|\bx,\mathcal{W})$ has the properties of a valid CDF. To satisfy these two requirements, we model $\theta_m(\bx,\mathcal{W})$ using a feed-forward neural network (FNN) with softmax output activation,
\begin{equation*}
   \begin{split}
     \theta_m(\bx,\mathcal{W}) =& \frac{\exp\{u_m(\bx,\mathcal{W})\}}{\sum_{i=1}^{r+p-1}\exp\{u_i(\bx,\mathcal{W})\}}\\
     u_m(\bx,\mathcal{W}) =& W_{2m0} + \sum_{l=1}^{V}W_{2ml}\phi\lpth W_{1l0} + \sum_{j=1}^dW_{1lj}x_j\rpth
   \end{split},
\end{equation*}
where $\mathcal{W} = \{W_{uvw}\}$ are the unknown weights and $\phi$ is the known activation function. Throughout this paper, $\phi$ is taken to be the hyperbolic tangent function. Profiting from its universal approximation theorem \citep{hornik1989}, FNN allows the unconstrained coefficient functions $u_m(\bx,\mathcal{W})$ to describe arbitrarly complex covariate effects. While the softmax activation naturally projects  $u_m(\bx,\mathcal{W})$ to the unit simplex and overcomes the challenge of parameter estimation under monotonicity constraints. For simplicity, we describe the FNN with a single hidden layer with $V$ neurons, but extensions to deeper networks are straightforward.

The proposed model can approximate any continuous conditional CDF. Following the results of \citet{chui1980} and \citet{abrahamowicz1992}, with a large enough $p$, we can assume for any $\bx$ there exists a set of non-negative coefficients $\{\alpha_1,\alpha_2,...,\alpha_{r+p+1}\}$ satisfying the constraint $\sum_{m=1}^{r+p+1}\alpha_{m}=1$ such that $\sum_{m=1}^{r+p+1}\alpha_mI_{m,r}(z|\bT)$ approximates the conditional CDF $F(z|\bx)$ arbitrarily well. In QUINN, the mapping $\psi:\bx\rightarrow\{\alpha_1,\alpha_2,...,\alpha_{r+p+1}\}$ is modeled by a single-hidden-layer FNN with softmax output which is a non-constant, bounded, and continuous function. Then by the universal approximation theorem \citep{hornik1989}, there exist weights $\mathcal{W}$ such that the single-hidden-layer FNN $\theta_m(\bx,\mathcal{W})$ approximates the mapping $\psi$ arbitrarily well for all $\bx$, provided that the number of hidden neurons $V$ is large enough. Thus, by leveraging the approximation power of I-splines and FNN, the model $\sum_{m=1}^{r+p+1}\theta_m(\bx,\mathcal{W})I_{m,r}(z|\bT)$ can approximate any conditional CDF $F(z|\bx)$.

We adopt a Bayesian framework to estimate the weights $\mathcal{W}$ by assigning them prior distributions. Compared to its frequentist counterpart, Bayesian neural network modeling can capture uncertainty in both the fitted model and weight parameters, and avoid over-fitting when the sample size is small. Zero-mean Gaussian distributions are the most commonly used prior on weights and have been explored in many classic works \citep{mackay1992,neal1993}. Their popularity arise from their ``weight-decay" regularization effect that prevents individual nodes from having extreme value. For QUINN, we set $W_{1vw}\indep\mathcal{N}(0,\sigma_w^2)$, $W_{2vw}\indep\mathcal{N}(0,\gamma^2)$ so that weights in input-hidden layer have feature-wise variances, and weights in hidden-output layer share a common variance. The scale hyperparameters $\sigma_w$ and $\gamma$ are also treated as unknown and assigned hyperpriors, so that their values can be optimized by the data. \citet{gelman2006} recommends half-$t$ families with a small degrees of freedom. These distributions allow the variance to be arbitrarily close to 0 which regularizes the complexity of the model. In practice however, the heavy-tailedness of half-$t$ distributions make them too broad and often cause difficulty in convergence. Therefore we set $\sigma_{w},\gamma\iid\mathcal{N}^+(0,a^2)$ to follow the half-Gaussian distribution for simpler posterior geometry. The variance of half-Gaussian prior is set to be $a^2=900$ so it is still relatively noninformative. Experiments show that our model is not sensitive to moderately large values of $a$. The likelihood function of QUINN has a closed-form expression, therefore MCMC algorithms can be used to explore the posterior. However, traditional methods such as random-walk Metropolis and Gibbs sampler do not scale well to high-dimensional posterior with complex geometry. In this paper, we use No-U-Turn sampler (NUTS) \citep{hoffman2014} that uses gradient information to sample efficiently from high-dimensional posterior. Appendix A describes the MCMC algorithm used to approximate the posterior.

Our ultimate goal is to estimate the quantile process of the original response variable $Q_Y(\tau|\bx)$. Let $\hat{F}_Z(z|\bx)$ denote the conditional CDF estimator and $\bD_{Z}=\{\tilde{z}_1,\tilde{z}_2,\dots,\tilde{z}_N\}$ denote a dense grid on the unit interval. Non-parametric estimate of the quantile process $Q_Z(\tau|\bx)$ can be easily obtained by first evaluating $\hat{F}_Z(z|\bx)$ on $\bD_Z$ and then performing linear interpolation on a dense percentile grid by treating $\{\hat{F}_Z(z_i|\bx)\}_{i=1}^N$ as the input values and $\bD_Z$ as the functional output values. Because of the one-to-one correspondence between quantile function and CDF, the resulting quantile process estimator will also inherit the approximator property of the proposed CDF estimator. Finally, the estimated quantile process of the original response is given by $\hat{Q}_Y(\tau|\bx) = g^{-1}\lsbk \hat{Q}_Z(\tau|\bx)\rsbk$.  

As discussed in Section 1, the proposed model has several advantages over existing non-linear SQR models  \citep{IzbickiLee2016,Cannon2018,DasGhoshal2018}. The combination of I-splines and FNN leads to a valid probability model that spans a wide class of conditional distribution functions.  Also, as described below and shown later by simulation studies, this combination leads to efficient computation and fully-Bayesian inference on quantile effects.
 
\section{Summarizing covariate effects}\label{s:sens}
QUINN includes a flexible FNN model for covariate effects across quantile levels. FNN is a ``black box" supervised learning model that excel in flexibility but lack transparency. Unlike linear QR models which enable rate-of-change interpretation of the $\tau$-dependent coefficients, the proposed FNN-based model does not characterize the covariate effects on the predicted quantile in a self-explanatory way. This is inconvenient, since QR models are often used for data exploratory purposes. 

Fortunately, research on model agnostic methods has allowed post hoc analysis of main effects and second-order interaction effects (we omit consideration of higher order effects as they cannot be visualized or interpreted meaningfully) of the covariates on the predictions made by ``black box" models. The most popular model agnostic method is partial dependence plot (PD plot) which visualizes the average marginal effect a (pair of) covariate(s) have on the predictions. The PD plot is straightforward to implement and intuitive to interpret, but is expensive to compute.  Recently, \citet{ApleyZhu2020} proposed accumulative local effects (ALEs) plot that provides the same level of interpretation in a more computationally efficient way. In this section, we provide a brief review of the definitions of ALE plot and explain how it can be applied to QR models. We will later demonstrate using a multivariate simulation study how ALE plot can be utilized to extract interpretation from QUINN. 

The sensitivity of $Q_Y(\tau|\bx)$ to covariate $j$ is naturally quantified by the derivative $q_j(\tau,\bx) = \partial Q_Y(\tau|\bx)/\partial x_j$.  In a linear QR, the derivative is the scalar effect of covariate $j$ on quantile level $\tau$, but for a non-linear regression function the derivative depends on $\bx$.  The ALE begins by averaging $q_j(\tau,\bX)$ over $\bX$ conditioned on $X_j = x_j$, giving ${\bar q}_j(\tau,x_j) = \mbox{E}_{\bX}\lsbk q_j(\tau,\bX)|X_j=x_j\rsbk$. The uncentered ALE main effect function of $X_j$ is then defined as 
\begin{equation*}
    {\bar Q}_j^U(\tau,x_j) =\int_{x_{\min,j}}^{x_j}{\bar q}_j(\tau,u_j)du_j.
\end{equation*}
The function ${\bar Q}_j^U(\tau,x_j)$ can be interpreted as the ALE of $X_j$ in the sense that it is an accumulation of local effects ${\bar q}_j(\tau,u_j)$ averaged over the distribution of $\bX$. The uncentered ALE effect does not have a straightforward interpretation because the derivative is invariant to scalar addition, which leads to the definition of the (centered) ALE main effect function ${\bar Q}_j(\tau,x_j)$ that is the same as ${\bar Q}_j^U(\tau,x_j)$ except centered to have mean 0 with respect to the marginal distribution of $X_j$. 

Analogous formulas define the second-order ALE for $X_j$ and $X_l$.  Consider the second-order partial derivative $q_{jl}(\tau,\bx) = \partial^2 Q_Y(\tau|\bx)/\partial x_j\partial x_l$.  The local effect at $X_j=x_j$ and $X_l=x_l$, averaging over the other covariates, is ${\bar q}_{jl}(\tau,x_j,x_l) = \mbox{E}_{\bX}\lsbk q_{jl}(\tau,\bX)|X_j=x_j,X_l=x_l\rsbk$. The uncentered second-order ALE is then
\begin{equation*}
    {\bar Q}_{jl}^U(\tau,x_j,x_l) =\int_{x_{\min,j}}^{x_j}\int_{x_{\min,l}}^{x_l}{\bar q}_{jl}(\tau,u_j,u_l)du_jdu_l,
\end{equation*}
and the second-order ALE function ${\bar Q}_{jl}(\tau,x_j,x_l)$ is mean-centered with respect to the marginal distribution of $(X_j,X_l)$. The second-order ALE ${\bar Q}_{jl}(\tau,x_j,x_l)$ describes the joint effects of the two covariates, which consist of both their main effects and interaction effect. In cases where assessment of only the interaction effect is of interest, main effects of $X_j$ and $X_l$ can be further subtracted from ${\bar Q}_{jl}(\tau,x_j,x_l)$ to obtain the pure interaction effect ${\bar Q}_{jl}^I(\tau,x_j,x_l)$.

The functions ${\bar Q}_{j}(\tau,x_j)$, ${\bar Q}_{jl}(\tau,x_j,x_l)$, and ${\bar Q}_{jl}^I(\tau,x_j,x_l)$ can be plotted to understand each main and interaction effect. When plotted against $x_j$, the main ALE $\bar{Q}_{j}(\tau,x_j)$ quantifies the difference between average prediction conditioned on $X_j=x_j$ and the average prediction over $\bX$.  When plotted against $x_j$ and $x_l$, the second-order ALE ${\bar Q}_{jl}(\tau,x_j,x_l)$ quantifies the difference between average prediction conditioned on $(X_j,X_l)=(x_j,x_l)$ and the average prediction over $\bX$. The interaction ALE ${\bar Q}_{jl}^I(\tau,x_j,x_l)$ can be interpreted analogously to ${\bar Q}_{jl}(\tau,x_j,x_l)$, except now the difference is contributed entirely by the interaction effect.

It is also useful to summarize the main and interaction ALEs with a one-number summary that can be used to rank the importance of each effect. Following \cite{greenwell2018}, we propose to measure overall variable importance (VI) for continuous covariates using the standard deviation of the ALE with respect to the marginal distribution of $\bX$, i.e., $\mbox{VI}_j(\tau) = \mbox{SD}\lsbk {\bar Q}_j(\tau,X_j)\rsbk$ and $ \mbox{VI}_{jl}(\tau) = \mbox{SD}\lsbk {\bar Q}^I_{jl}(\tau,X_j,X_l)\rsbk$. For categorical covariates, the standard deviation is replaced by one fourth of the range. These VI scores (and the intermediate functions ${\bar Q}_j$, ${\bar Q}_{jl}$, and ${\bar Q}^I_{jl}$) can be approximated using the partitioning schemes of \cite{ApleyZhu2020} as described in Appendix B.  

Although for notational simplicity we have omitted the dependence of the quantile function on the parameters $\mathcal{W}$, in practice the posterior uncertainty in $\mathcal{W}$ leads to posterior uncertainty in the sensitivity metrics such as ${\bar Q}_j(\tau)$ and $\mbox{VI}_j(\tau)$.  We account for this uncertainty by computing the sensitivity measures for many MCMC samples from the posterior distribution of $\mathcal{W}$, giving a Monte Carlo approximation of the posterior distribution of the sensitivity measures.   

\section{Simulation}\label{s:simu}
We investigate the numerical performance of our model in four scenarios. The details of each simulation design are provided below.

\begin{DS}
\item  The covariate and response are generated as $X\sim\mbox{Uniform}(0,5)$ and
\begin{equation*}
   Y = X + \sin(2X) + 3\epsilon;\ \epsilon\sim\mbox{Skew-Normal}(0,1,4).
\end{equation*}
The quantile curves are parallel, and the data exhibit strong right-skewness.
\item The covariate and response are generated as $X\sim\mbox{Uniform}(0,1)$ and
\begin{equation*}
Y = 3X + [0.5+2X+\sin(3\pi X+1)]\epsilon;\ \epsilon\sim\mbox{Normal}(0,1).
\end{equation*}
The data exhibit strong heteroscedasticity. The quantile curves are linear at the median but have strong curvature at the extremes.
\item  The covariates $X_j,\ j=1,2$ are generated from $\mbox{Uniform}([0,1]\times[0,1])$. The response variable $Y$ is given by
\begin{equation*}
            Y=\sin(2\pi X_1) + \cos(2\pi X_2) + \sqrt{2(X_1^2+X_2^2)}\epsilon;\ \epsilon\sim\mbox{Student's } t(3).
\end{equation*}
The data exhibit both heteroscadasticity and heavy-tailedness.
\item The covariates $X_j,\ j=1,2,..,d$ are generated from $\mbox{Uniform}([0,1]^{d})$. The quantile function $Q_Y(\tau|\bX)$ is given by
\begin{equation*}
\begin{split}
            Q_Y(\tau|\bX) =&\ 3(\tau-0.5)\lpth X_1+\frac{3}{5}\rpth^3\\
            &+15\lsbk X_2+4\lpth X_2-\frac{1}{2}\rpth^2\rsbk\exp\lpth -X_2^2\rpth\\
            &+12\exp\lsbk \lpth X_3+\frac{1}{2}\rpth^2\lpth X_4-\frac{1}{2}\rpth^2\rsbk\\
            &+5(\tau-1)\lpth X_5+\frac{2}{5}\rpth\lpth X_6+\frac{1}{2}\rpth^2
            +0.25\Phi^{-1}(\tau),
\end{split}
\end{equation*}
where $\Phi^{-1}(\cdot)$ is the standard normal quantile function and $d\in\{10,20,40\}$. The response variable is generated by sampling $U\sim\mbox{Uniform}(0,1)$ and setting $Y=Q(U|\bX)$. The quantile process has a complex structure with strong interaction effects. The model is sparse as only the first six covariates affect the quantile function.
\end{DS}

\begin{figure}
\centerline{\includegraphics[page=2,width=\linewidth]{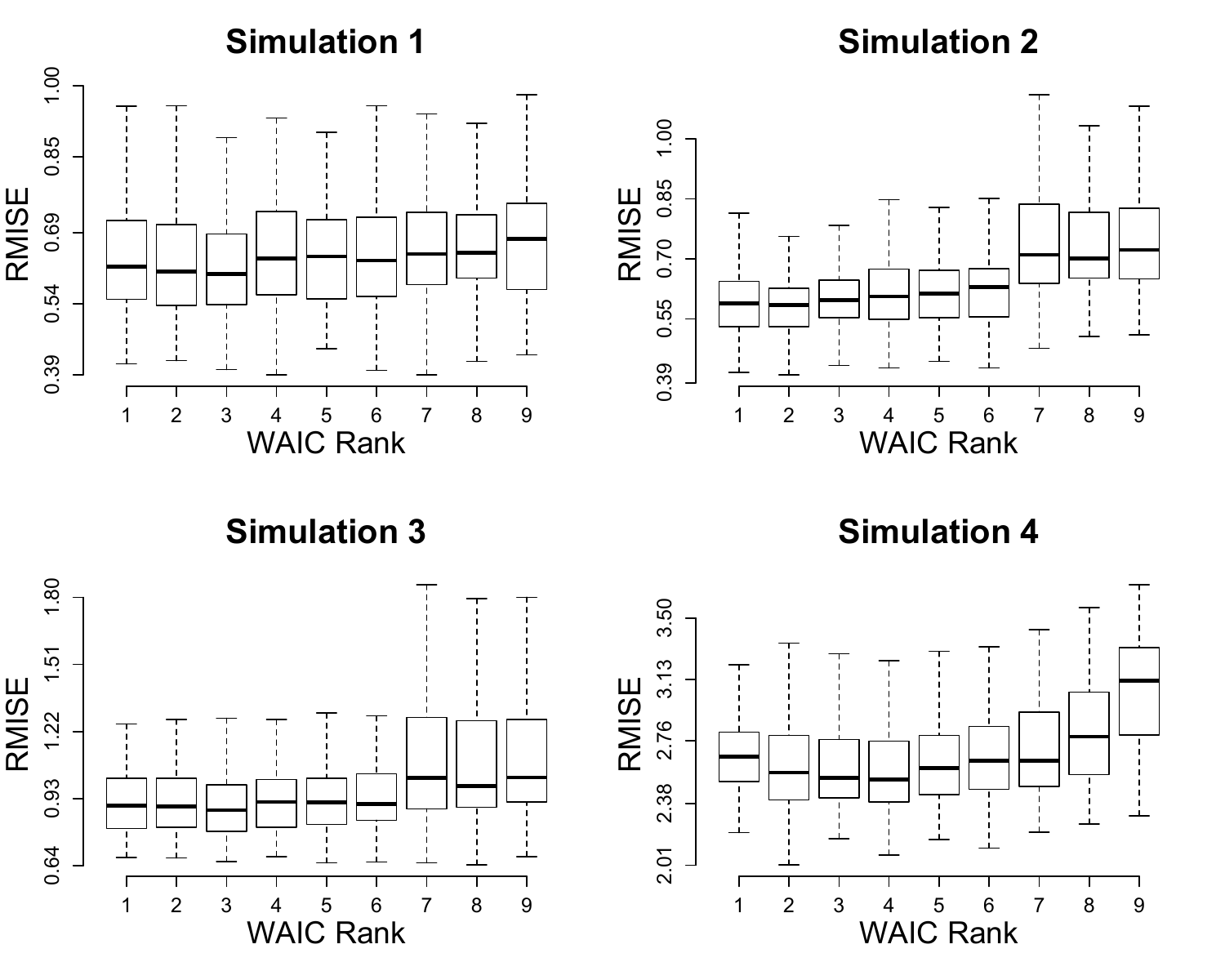}}
\caption{Distribution of RMISE conditioned on rank of WAIC, constructed using 100 replicates of each simulation study.}\label{f:waic}
\end{figure}

For Designs 1--3, we generate samples of sizes $n\in\{50, 100, 200\}$ and for Design 4 we use $n=200$. The proposed model is compared to four non-linear SQR methods: the monotone composite QR neural network (MCQRNN) of \citet{Cannon2018}, the non-parametric simultaneous QR (NPSQR) of \citet{DasGhoshal2018}, the non-parametric distribution function simultaneous QR (NPDFSQR) also of \citet{DasGhoshal2018}, and the spectral series conditional density estimator (seriesCDE) of \citet{IzbickiLee2016}. MCQRNN is implemented in the \texttt{qrnn} package in R; codes for NPSQR and NPDFSQR are available from the second author's webpage; and codes for seriesCDE are available from the supplemental material of their online paper. Implementation details including model selection for the competing methods are given in Appendix C. For QUINN, we first map the response variable to the unit interval using min-max normalization. The covariates are not required to be normalized. However, it is a common practice to normalize the inputs to a FNN when optimizing its parameters using a gradient-based approach \citep{bishop1995}. In this paper, we always map the covariate vector to the unit interval, even if it is one-dimensional. Posterior distribution of QUINN is approximated by 1900 MCMC samples that are obtained by running NUTS for 20,000 iterations, discarding the first 1000 iterations as burn-in and saving every 10th draw from the remaining iterations. Convergence of MCMC is monitored by trace plots of log-likelihood from multiple independent chains as shown in Figure~\ref{f:trace}, and popular diagnostic statistics as described in Appendix A.6. The performance of QUINN depends on the number of spline knots $p$ and hidden neurons $V$, so we use a grid search approach and select the best combination of $p, V\in\{5,8,10\}$ based on WAIC \citep{watanabe2013}. We choose WAIC over other information criteria (e.g. AIC and DIC) because it is fully Bayesian, uses the entire posterior distribution, and is asymptotically equal to Bayesian leave-one-out cross-validation \citep{vehtari2017}. Figure~\ref{f:waic} plots the distribution of out-of-sample RMISE against ranking of WAIC. The result shows that model chosen by WAIC is in favor of a higher out-of-sample prediction accuracy. We also observe that the performance of QUINN is generally robust to different values of the two parameters except for some particularly bad combinations.

\begin{figure}[b]
\centerline{\includegraphics[page=1,width=\linewidth]{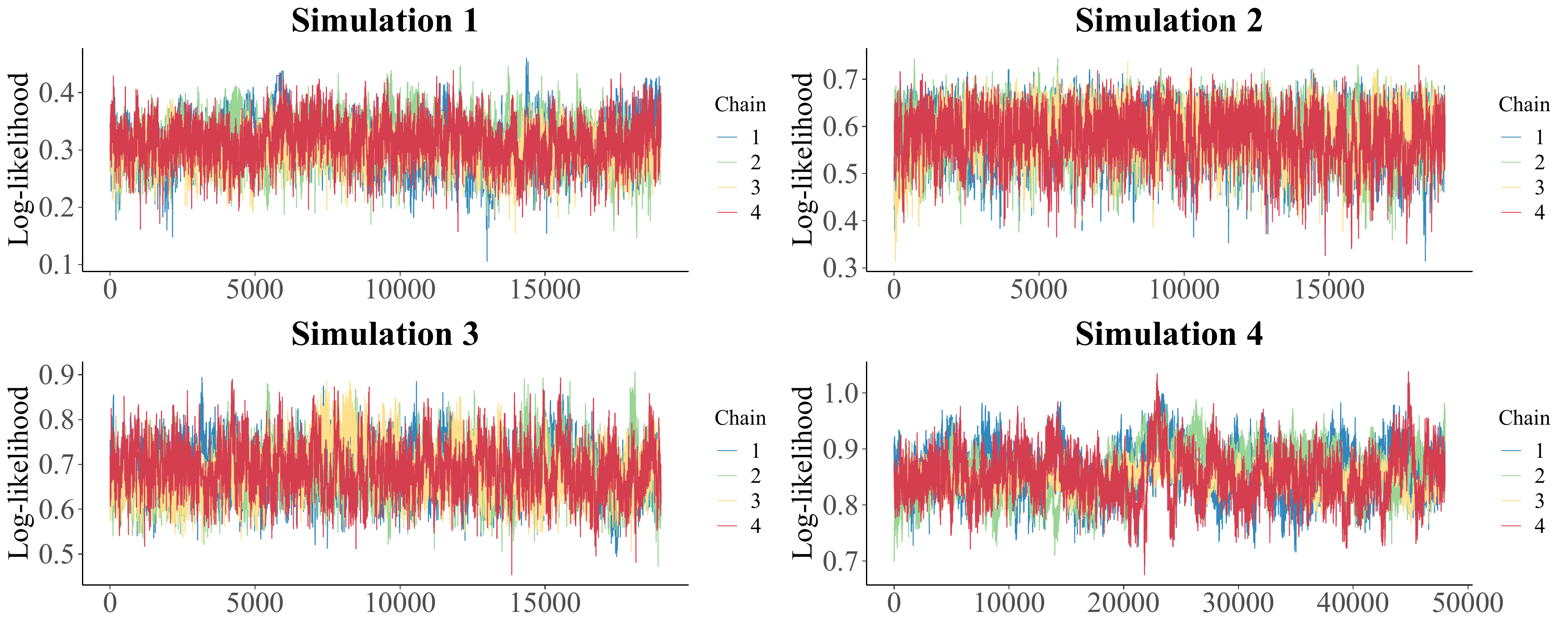}}
\caption{Trace plot of log-likelihood showing convergence and good mixing of MCMC chains.}\label{f:trace}
\end{figure}

To compare the different approaches, 100 data sets are simulated. For each sample size, the performance of each method is measured by the root mean integrated square error (RMISE) between the actual and estimated (posterior mean) quantile processes. We first divide the domain of each dimension of $\bX$ by $g$ equidistant grid-points, giving $G=gd$ vectors $\tilde{\bx}_1,...,\tilde{\bx}_G$ that span the range of $\bX$. For Designs 1 and 2, we set $g=G=101$; for Design~3, we set $g=21$ and thus $G=21^2$. The RMISE is then approximated as
\begin{equation*}
        \mbox{RMISE}(\tau_k)= \sqrt{\frac{1}{G}\sum_{i=1}^{G} \lcbk Q_Y\lpth \tau_k|\tilde{\bx}_i\rpth-\hat{Q}\lpth \tau_k,\tilde{\bx}_i\rpth\rcbk^2}
\end{equation*}
for quantile level $\tau_k\in\{0.05,0.10,...,0.95\}$ and 
\begin{equation*}
        \mbox{RMISE}_{\text{QP}} = \sqrt{\frac{1}{19}\sum_{k=1}^{19}\mbox{RMISE}(\tau_k)^2}
\end{equation*}
for the entire quantile process.

\begin{figure}
\centerline{\includegraphics[width=\linewidth]{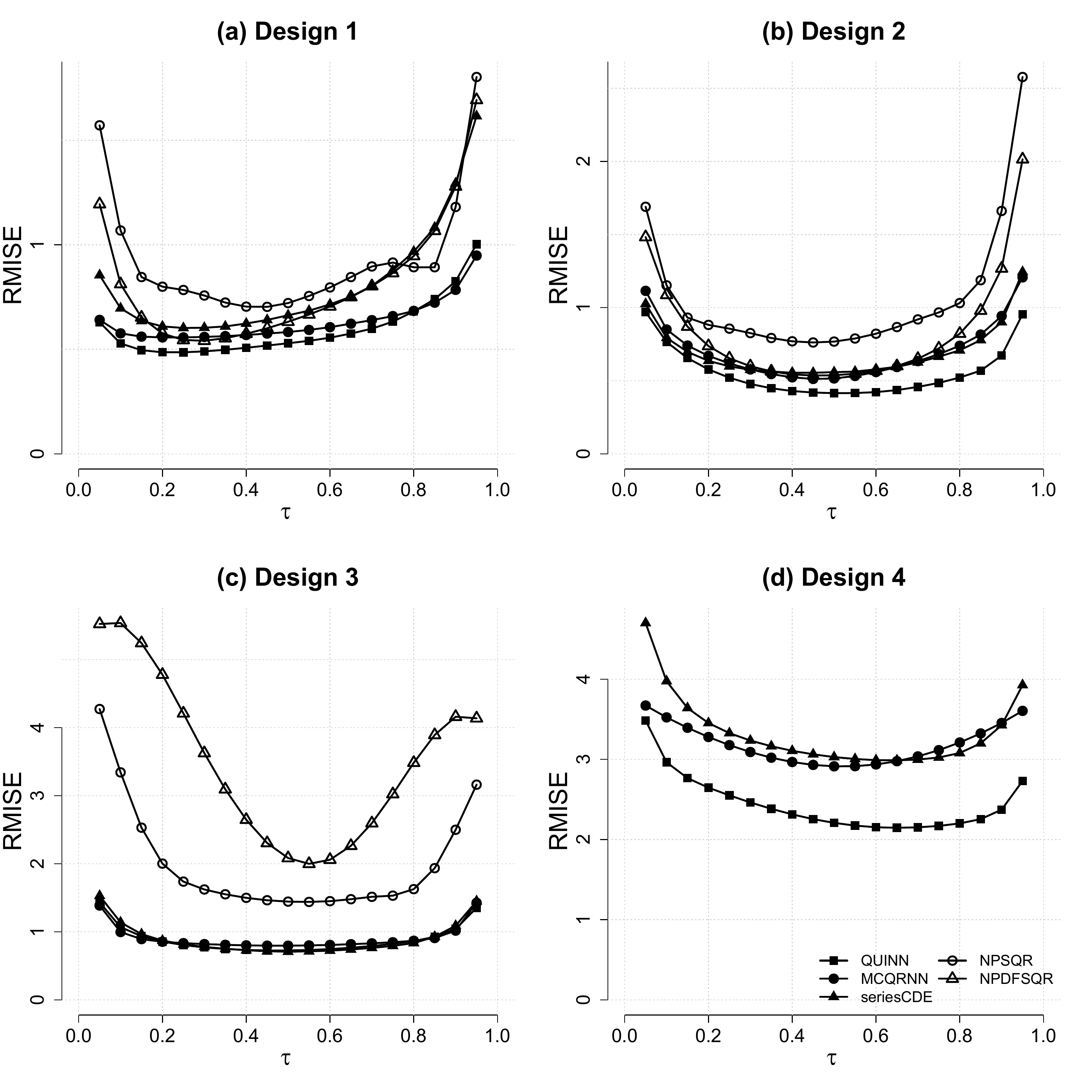}}
\caption{Average $\mbox{RMISE}(\tau)$, $\tau\in\{0.05,0.1,...,0.95\}$ over $100$ simulated data sets by quantile level $\tau$ for the simulation study. The training sample size is 100 for Designs 1--3, and 200 for Design 4.}\label{f:sim_rmise}
\end{figure}

The average $\mbox{RMISE}_{\text{QP}}$ over $100$ simulated data sets along with their standard errors are shown in Table~1. The results show that QUINN yields significantly smaller average $\mbox{RMISE}_{\text{QP}}$ in all settings when compared to NPSQR, NPDFSQR, and seriesCDE; and smaller or similar average $\mbox{RMISE}_{\text{QP}}$ in all but one setting when compared to MCQRNN. In particular, QUINN is robust to data sparsity as seen from its small variance. We also plot the average $\mbox{RMISE}({\tau})$ for cases when $n=100$ in Figure~\ref{f:sim_rmise}(a)--(c). The results show that QUINN gives the best estimation of intermediate quantiles in all cases, whereas MCQRNN gives better estimation of extreme quantiles when the data exhibit significant heavy-tailedness. 

\begin{table}[h]
\begin{center}
\begin{minipage}{130mm}
\caption{Simulation results: Average $\mbox{RMISE}_{\text{QP}}$ over $100$ replicates with standard error in parentheses, and the smallest error in each row is in bold.}
\label{t:sim_rmise}
\end{minipage}
\begin{tabular}{c|c|ccccc}
\hline
\hline
Design & $n$ & QUINN & MCQRNN & NPSQR & NPDFSQR & seriesCDE\\ 
\hline
\multirow{3}{*}{1} & 50 &  \textbf{0.86} (0.22) & 1.11 (1.98) & 1.13 (0.26) & 1.15 (0.26)  &  1.11 (0.29)\\
 & 100 &  \textbf{0.62} (0.11) & 0.65 (0.15) & 1.00 (0.19) & 0.89 (0.17) &  0.87 (0.19)\\
 & 200 &  0.50 (0.09) & \textbf{0.47} (0.11) & 0.96 (0.18) & 0.74 (0.16) &  0.71 (0.19)\\
\multirow{3}{*}{2} & 50 &  \textbf{0.80} (0.16) & 1.18 (2.28) & 1.18 (0.28) & 1.20 (0.31)  & 0.90 (0.20)\\
 & 100 &  \textbf{0.60} (0.10) & 0.72 (0.13) & 1.19 (0.22) & 0.93 (0.19) &  0.74 (0.19)\\
 & 200 &  \textbf{0.48} (0.06) & 0.53 (0.11) & 1.03 (0.18) & 0.76 (0.15) &  0.57 (0.15)\\
\multirow{3}{*}{3} & 50 &  \textbf{1.19} (0.32) & 1.39 (0.81) & 2.23 (1.78) & 2.68 (2.25)  &  1.23 (0.53)\\
 & 100 &  \textbf{0.93} (0.21) & \textbf{0.94} (0.21) & 2.33 (1.83) & 4.04 (2.62) &  0.96 (0.27)\\
 & 200 &  0.82 (0.27) & \textbf{0.71} (0.34) & 2.16 (1.20) & 3.66 (2.29) &  0.86 (0.27)\\
 4 & 200 & \textbf{2.47} (0.25) & 3.21 (1.58) & - & - & 3.37 (0.21)\\
\hline
\hline
\end{tabular}
\end{center}
\end{table}

For Design 4, we compare QUINN with MCQRNN and seriesCDE only. We omit NPSQR and NPDFSQR because expanding each covariate of a $d$-dimensional $\bX$ using quadratic B-spline basis functions results in a parameter space of dimension $p_{\text{DG}}(p_{\text{DG}}+2)^d$. Even with $d$ as few as $10$, fitting NPSQR and NPDFSQR becomes computationally infeasible.
We generate sample of size $n=400$ and split into 200 training and 200 testing data points. Each model is first fit to the training data, then conditional quantile predictions at $\tau \in\{ 0.05,0.1,...,0.95\}$ are calculated for each given $\bx$ of the testing data. Posterior distribution of QUINN is approximated by 2400 samples that are obtained by running NUTS for 50,000 iterations, discarding the first 2000 iterations as burn-in and saving every 20th draw from the remaining iterations. We choose the best model configuration of $p\in\{5,8,10\}$ and $V\in\{8,10,15\}$ using WAIC. We generate 100 replicates and compare different approaches based on $\mbox{RMISE}(\tau)$ and $\mbox{RMISE}_{\text{QP}}$ between actual and predicted quantiles conditioned on the testing data points. The average $\mbox{RMISE}_{\text{QP}}$ for $d=10$ are 2.47 for QUINN, 3.21 for MCQRNN, and 3.37 for seriesCDE and the average $\mbox{RMISE}(\tau)$ are plotted in Figure~\ref{f:sim_rmise}(d). The result shows that QUINN gives substantially better estimation of the quantile process than MCQRNN and seriesCDE. To further investigate the performance of QUINN in high-dimension setting, we repeat Design 4 with $\bX$ of dimensions $d=20,40$ with the additional covariates being independent of the response. The average $\mbox{RMISE}_{\text{QP}}$ are 2.73 and 3.09, respectively. Therefore, QUINN shows promising performance when the quantile process is high-dimensional, has complex interaction effects, and has a sparse structure.

\begin{figure}
\centerline{\includegraphics[page=1,width=\linewidth]{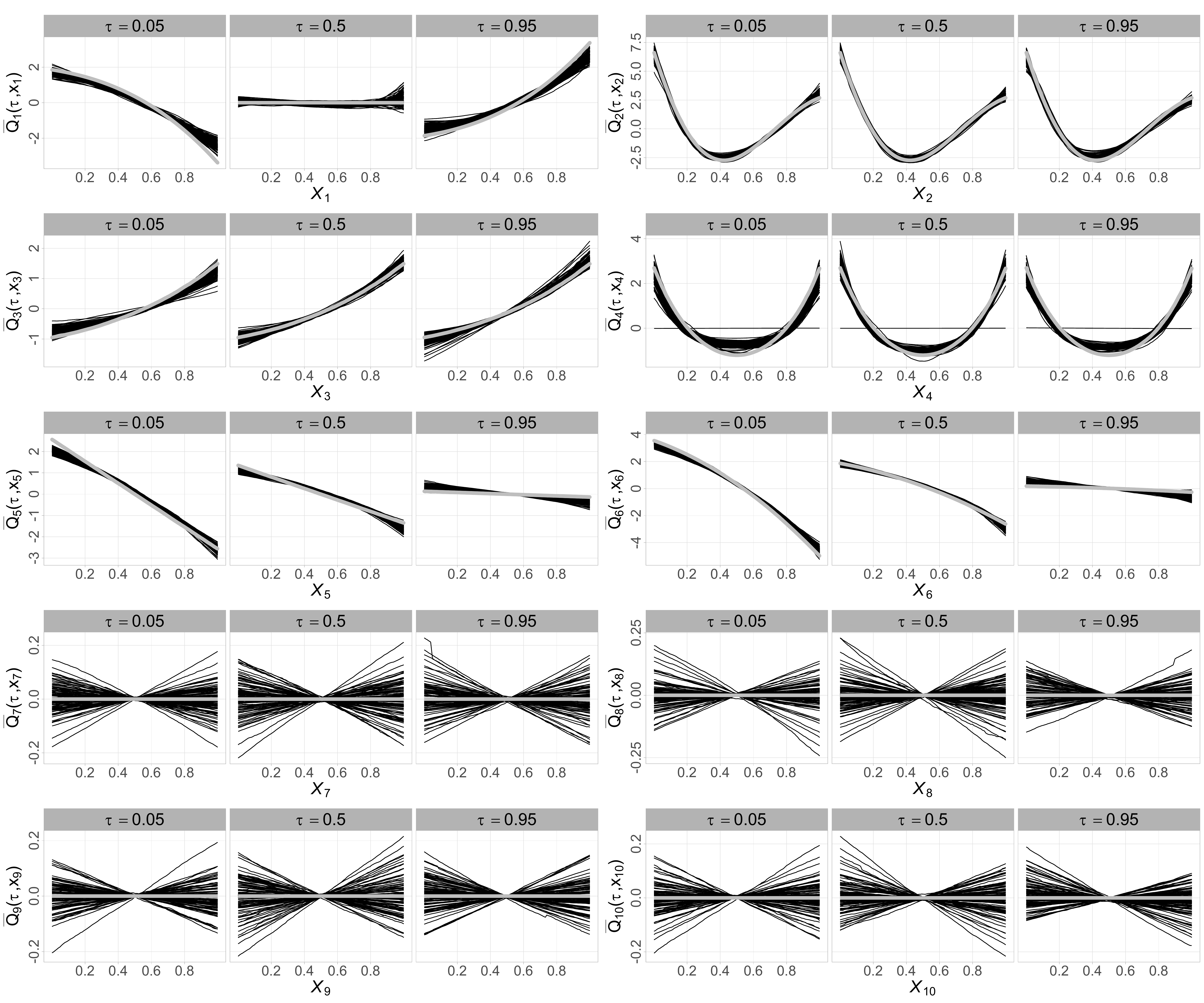}}
\caption{Marginal main effect estimates for Simulation Design~4. Accumulative local effects (ALE) ${\bar Q}_j(\tau,x_j)$ for $j\in\{1,...,10\}$ and $\tau \in\{ 0.05,0.50,0.95\}$. Black lines represent individual ALE calculated from the 100 replicates, and the light gray line represents the true ALE based on the generating model.}\label{f:sens_ale}
\end{figure}

We now demonstrate how ALEs plot can be used with QUINN to visualize main and second-order interaction effects on its predicted quantiles. Design~4 is constructed such that ${\bar Q}_{1}(\tau,x_1)$, ..., ${\bar Q}_{6}(\tau,x_6)$, ${\bar Q}_{3,4}(\tau,x_3,x_4)$, and ${\bar Q}_{5,6}(\tau,x_5,x_6)$ are non-zero functions of $\tau$. In addition, ${\bar Q}_{j}(\tau,x_j),$ $j\in\{1,5,6\}$ and ${\bar Q}_{5,6}(\tau,x_5,x_6)$ are non-constant functions of $\tau$. To evaluate the sensitivity of QUINN in identifying these marginal quantile effects, we generate 100 replicates of sample size 5000 from Design~4. For each replicate, we estimate the ALE main effect for each covariate and interaction effect for each pair of covariates at quantile levels $\tau\in\{0.05,0.10,...,0.95\}$ based on the fitted QUINN. The estimated ALE main effects at these quantile levels along with their ground truths are shown in Figure~\ref{f:sens_ale}, where the thin black lines represent individual estimates based on the 100 simulated data sets, and the thick gray line represents the true ALE effect calculated from the generating model. The estimated ALEs show that QUINN successfully captures the main effects of each covariate. For ALE interaction effects, since it is impossible to visualize all estimated surfaces in one plot, we instead show estimates from 8 randomly selected replicates. For each replicate, estimated ${\bar Q}_{jl}(\tau,x_j,x_l)$ and ${\bar Q}_{jl}^I(\tau,x_j,x_l)$ at quantile levels $\tau\in\{0.05,0.5,0.95\}$ are visualized using contour plots (see Appendix D) and compared with that of the ground truth. The results show that QUINN also successfully recovers the complex interaction effects of the generating model.

\begin{figure}
  \centerline{\includegraphics[page=2,width=\linewidth]{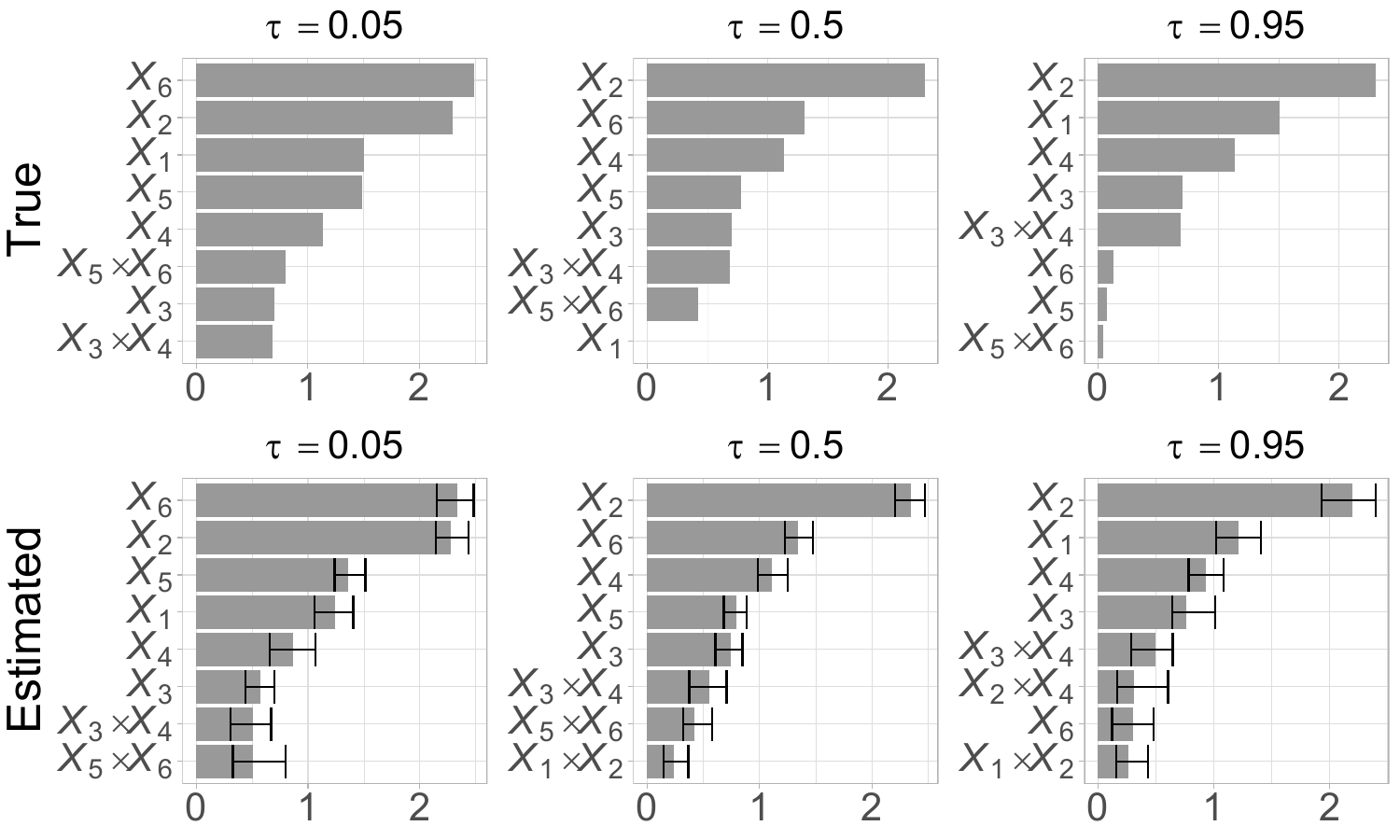}}
  \caption{Marginal effects importance for Simulation Design~4. Estimated and true variable importance for the top 8 marginal effects and $\tau\in\{0.05,0.50,0.95\}$; the estimates are averaged over 100 replicates and thin horizontal lines are 95\% credible intervals. }\label{f:sens_imp}
\end{figure}

To investigate whether QUINN is capable of recovering the relative importance of the marginal effects, we calculate VI scores for each ALE main and interaction effect. Because only eight marginal effects contribute to the conditional quantile function in Design~4, we demonstrate the sensitivity of QUINN by showing the rank plot of the top eight estimated marginal effects with the highest VI in Figure~\ref{f:sens_imp}. The results show that at quantile levels $\tau\in\{0.05, 0.5, 0.95\}$, the estimated VIs and their ranking of the top eight marginal effects resemble the ground truth. The sensitivity analysis shows that QUINN is able to identify the relative order of the important covariate effects.

% The estimated interaction effects are analyzed in Figure~\ref{fig:sens_int}, where the off-diagonal cells contain VI scores of pair-wise interaction effects. Our results show that when $\tau=0.05$ or $0.50$, the estimated VI scores of ${\bar Q}_{3,4}(\tau,x_3,x_4)$ and ${\bar Q}_{5,6}(\tau,x_5,x_6)$, the only two interaction effects that contribute to the generating model, are significantly higher than those of all other pairs; when $\tau=0.95$, ${\bar Q}_{5,6}(\tau,x_5,x_6)$ is close to zero and thus was difficult to distinguish from other non-contributing interaction effects. We further calculate the Monte Carlo probability that ${\bar Q}_{3,4}(\tau,x_3,x_4)$ or ${\bar Q}_{5,6}(\tau,x_5,x_6)$ is ranked top 2 among all interaction effects and show the results in Figure~\ref{fig:sens_int_rank}. The probability that ${\bar Q}_{3,4}(\tau,x_3,x_4)$ is ranked top 2 is close to $1$ across all quantiles; the probability that ${\bar Q}_{5,6}(\tau,x_5,x_6)$ is ranked top 2 is close to $1$ when $\tau < 0.6$, but decreases as ${\bar Q}_{5,6}(\tau,x_5,x_6)$ decreases. 

\section{Application to birth weight data}
\label{s:app}
%\label{s:app:birth}
To illustrate the practical effectiveness of QUINN, we study the effect of pregnancy-related factors on infant birth weight (\texttt{Weight}, in grams) quantiles. Our data consist of 10,000 randomly chosen entries from the 2019 U.S. Natality Data Set \citep{nchs2019} on singleton live births to mothers recorded as Black or White, in the age group 18--45, with height between 59 and 73 inches, and smoke no more than 20 cigarettes daily during pregnancy. The list of covariates contains demographic characteristics, maternal behavior and health characteristics, as well as infant health characteristics. For demographic characteristics, we include indicator of age above 40 years old (\texttt{fatherAge}) for the father; and age (\texttt{motherAge}, in years), indicators of Black (\texttt{Black}), education attainment up to high school graduate (\texttt{highSchool}) and at least college graduate (\texttt{collegeGraduate}), and parity greater than 1 (\texttt{Parity}) for the mother. For maternal behavior and health characteristics, we include body mass index (\texttt{BMI}), height (\texttt{Height}, in inches), weight gain (\texttt{wtGain}, in pounds), indicator of smoking before pregnancy (\texttt{Smoker}), average daily number of cigarettes during pregnancy (\texttt{Cigaretters}), indicators of not receiving prenatal care (\texttt{noPrenatal}), pre-existing diabetes (\texttt{preDiab}) and hypertension (\texttt{preHype}), gestational diabetes (\texttt{gestDiab}) and hypertension (\texttt{gestHype}), no infections present and/or treated during pregnancy (\texttt{noInfec}), and infertile treatment (\texttt{infTreat}). For infant health characteristics, we include gestational age (\texttt{Week}) and indicator of boy (\texttt{Boy}).

The response variable and all continuous covariates are mapped to the unit interval using min-max normalization. We fit QUINN with $V\in\{10,20,30\}$ hidden neurons and $p\in\{10,15,20\}$ spline knots. We approximate its posterior distribution using 2400 samples obtained by running NUTS for 50,000 iterations, discarding the first 2000 iterations as burn-in, and selecting every 20th draw from the remaining iterations. The best model configuration is chosen based on WAIC. 

\begin{figure}
%\centerline{\includegraphics[page=1,width=\linewidth]{figures/birthwt_analysis.pdf}}
\centerline{\includegraphics[page=1,width=0.85\linewidth]{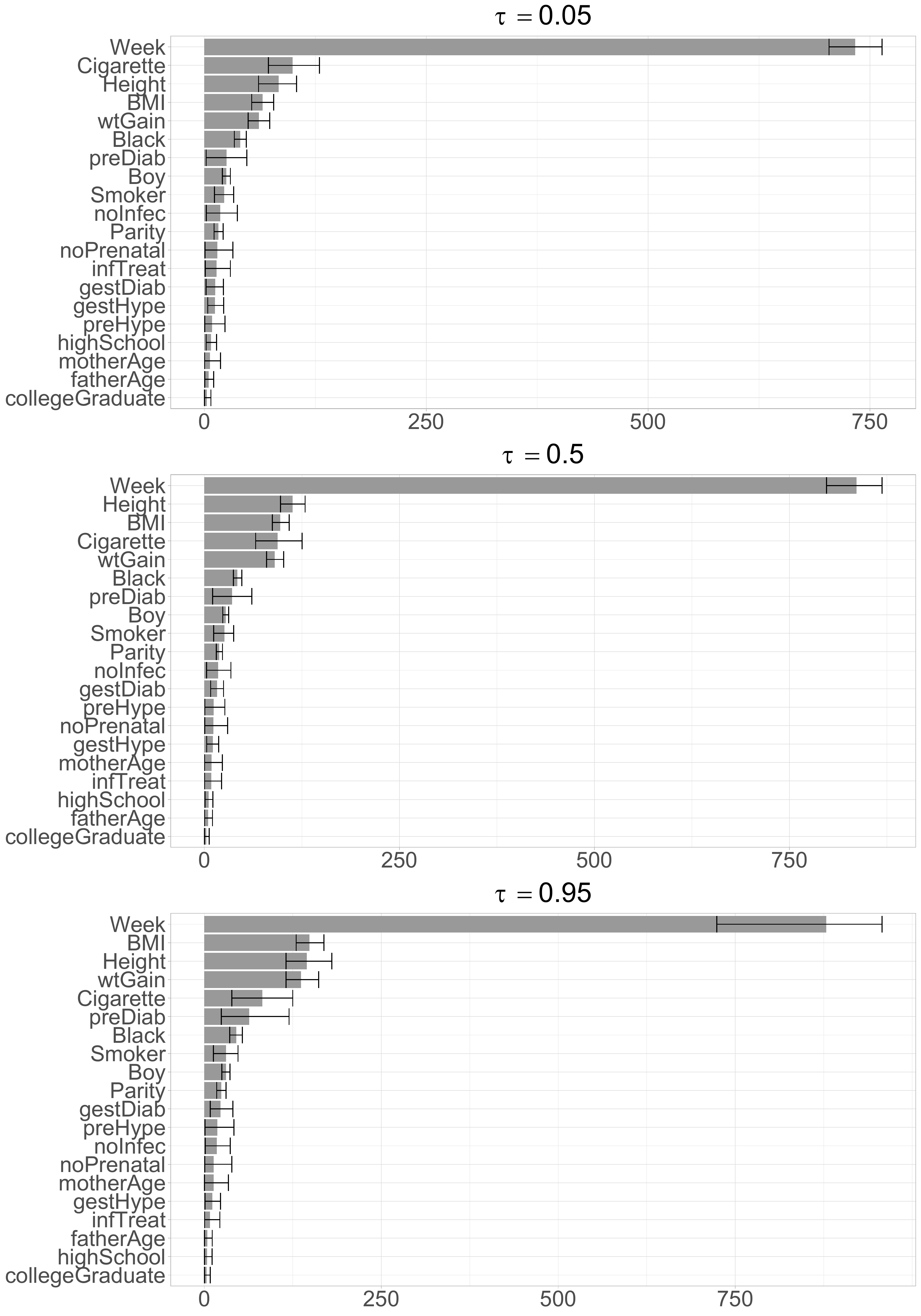}}
\caption{Posterior mean variable importance measure for all main effects at $\tau\in\{0.05,0.50,0.95\}$. The thin horizontal lines are 95\% credible intervals. }\label{f:birthwt_imp}
\end{figure}

To determine which covariates have the most significant impact on the birth weight quantiles, we calculate the ALE-induced VI score for each covariate across different quantile levels. In particular, we are interested in identifying the covariates that most impact LBW (represented by the 0.05 quantile), typical birth weight (TBW, represented by the 0.5 quantile), and HBW (represented by the 0.95 quantile). Figure~\ref{f:birthwt_imp} shows the ranking of ALE main effects at $\tau\in\{0.05,0.50,0.95\}$. The main effects of \texttt{Week}, \texttt{height}, \texttt{BMI}, \texttt{wtGain}, \texttt{Cigarette}, \texttt{Black}, \texttt{preDiab}, \texttt{Boy} and \texttt{Smoker} have the highest VI measure at all three quantiles and therefore are most influential on the birth weight distribution. In particular, \texttt{Week} has a dominant effect on all three quantiles, \texttt{Cigarette} is most influential on LBW, and \texttt{Height} and \texttt{BMI} are more influential on TBW and HBW.

\begin{figure}
\centerline{\includegraphics[page=2,width=0.25\textwidth]{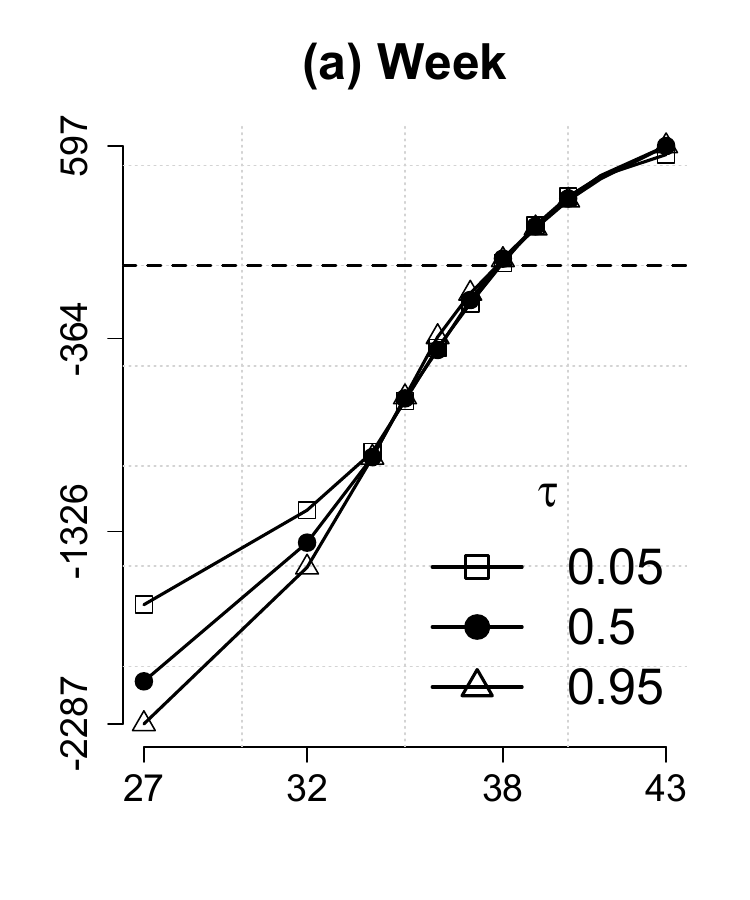}
\includegraphics[page=10,width=0.25\textwidth]{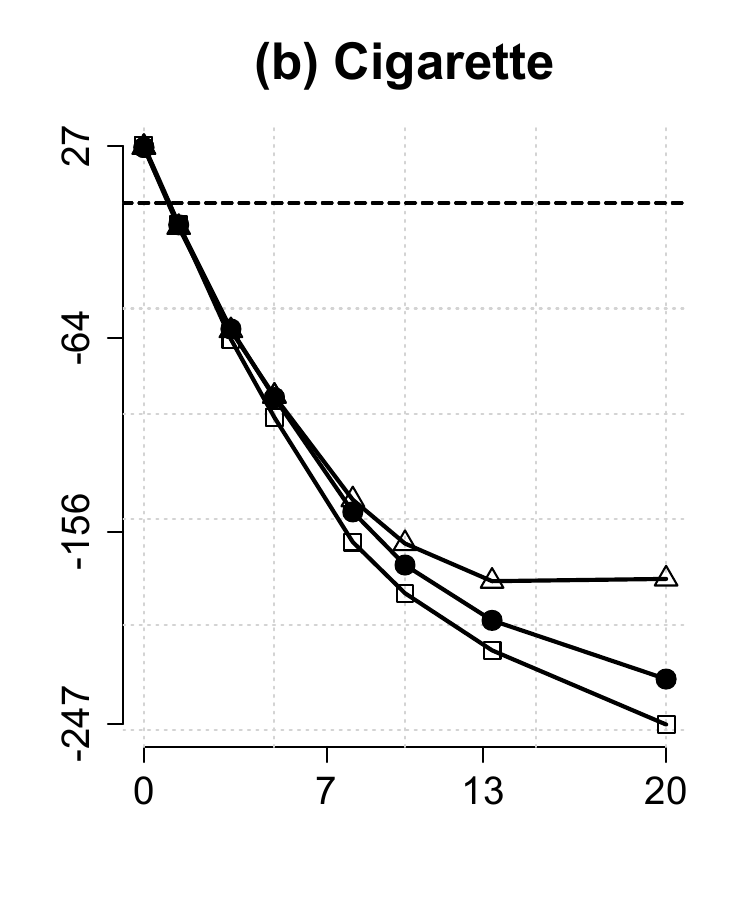}
\includegraphics[page=3,width=0.25\textwidth]{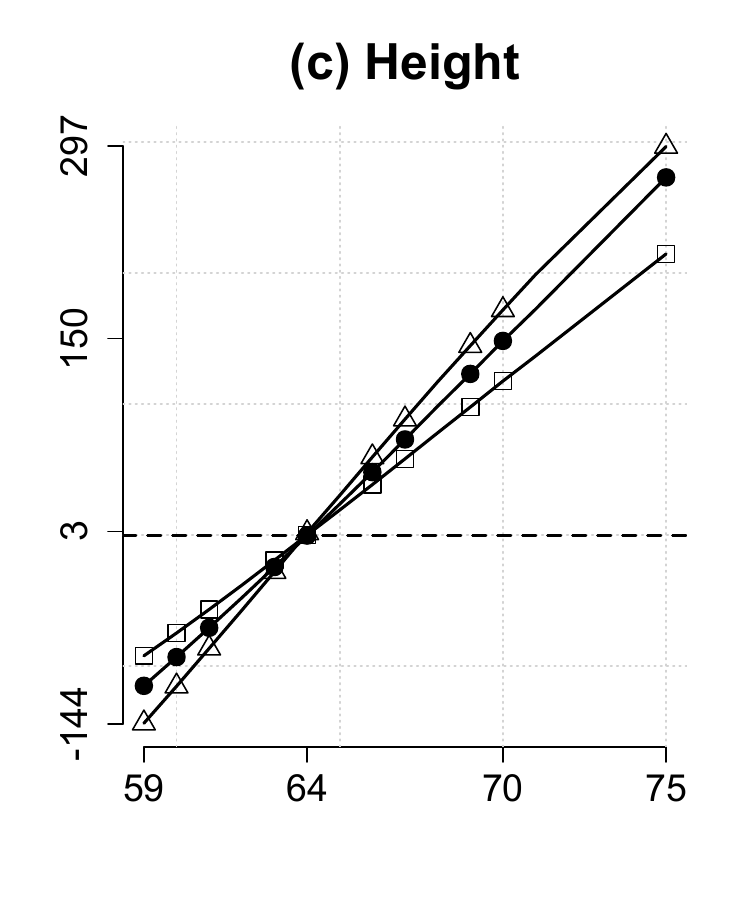}
\includegraphics[page=4,width=0.25\textwidth]{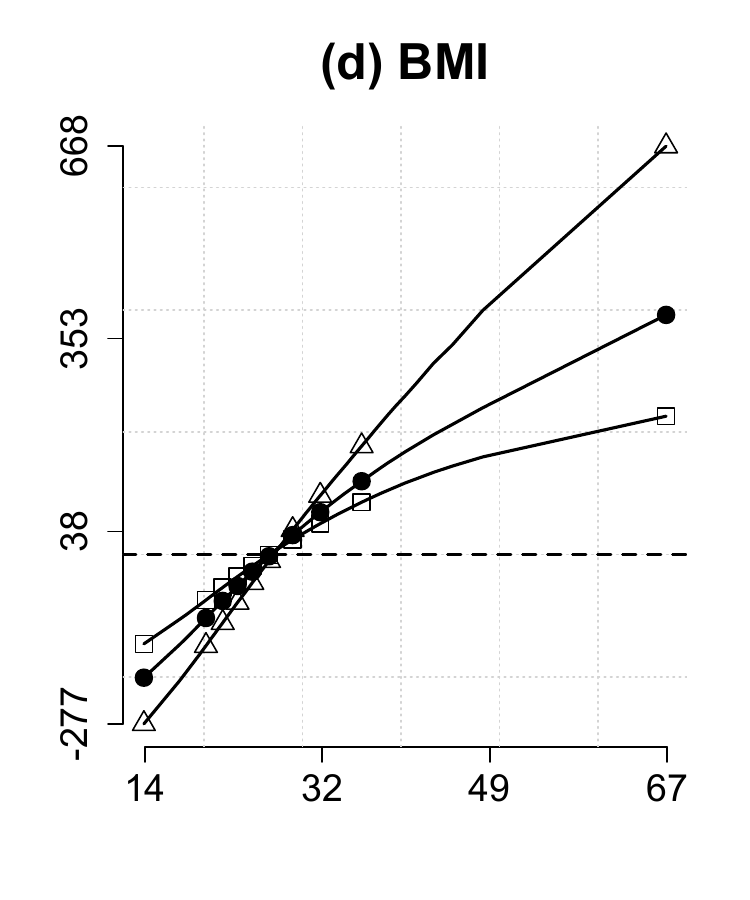}}
\centerline{
\includegraphics[page=5,width=0.25\textwidth]{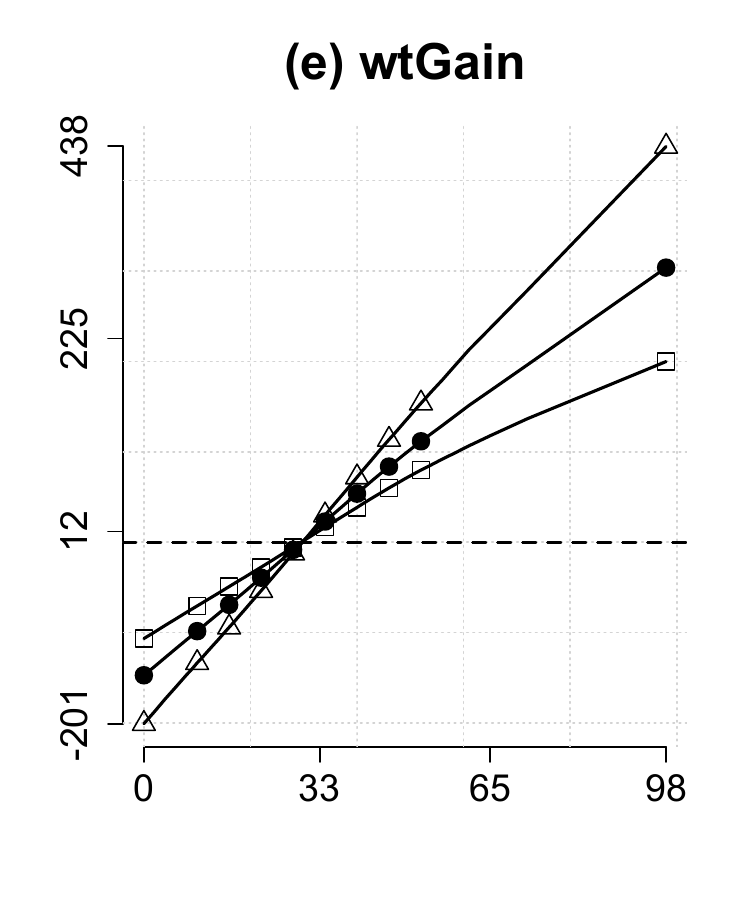}
\includegraphics[page=6,width=0.25\textwidth]{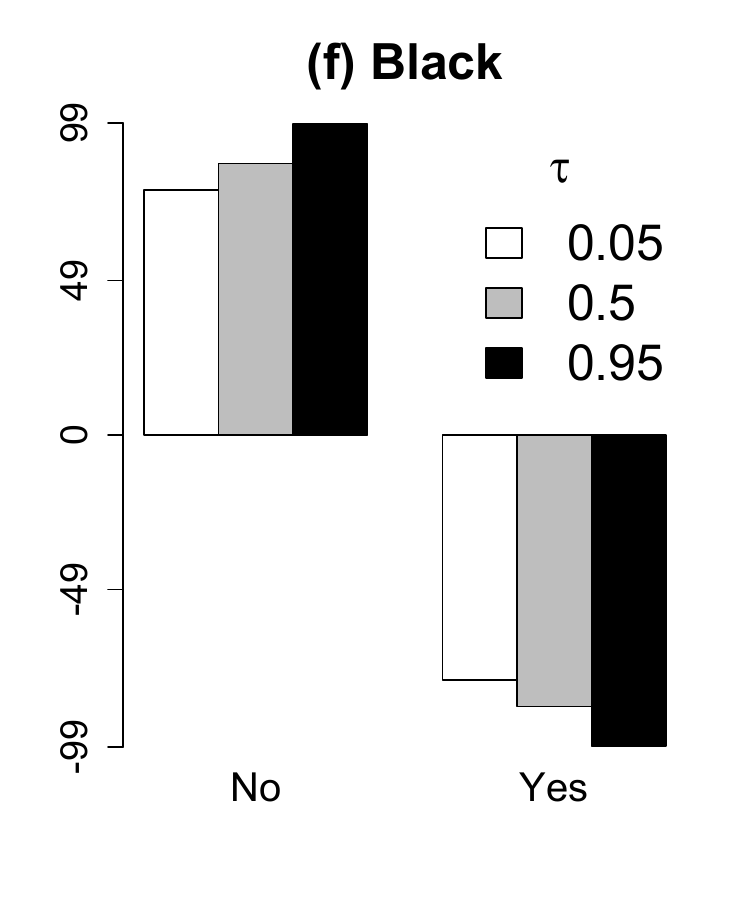}
\includegraphics[page=7,width=0.25\textwidth]{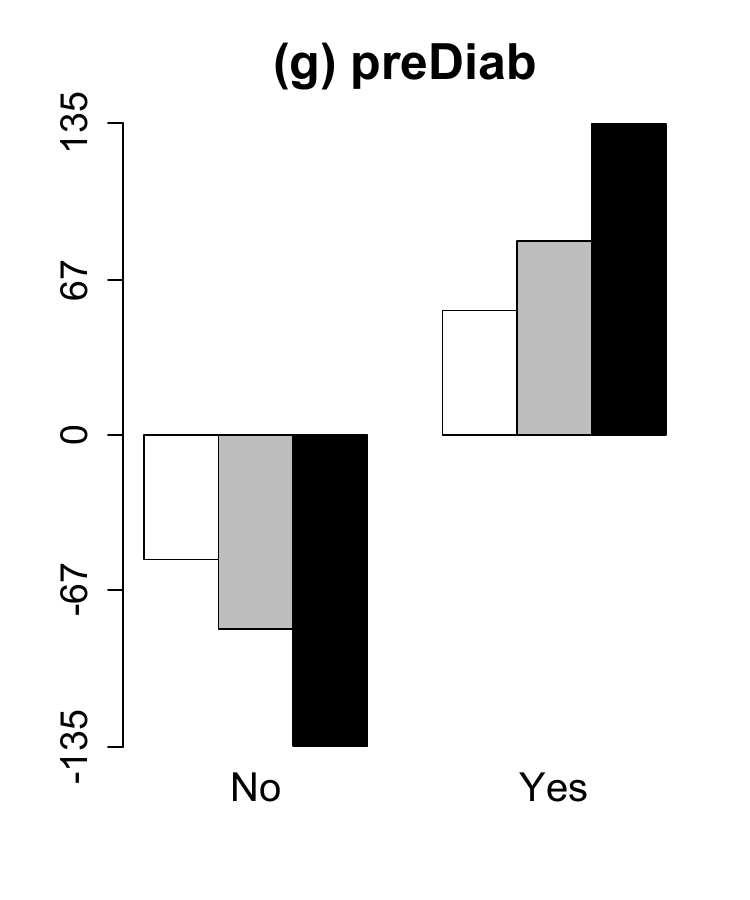}
\includegraphics[page=8,width=0.25\textwidth]{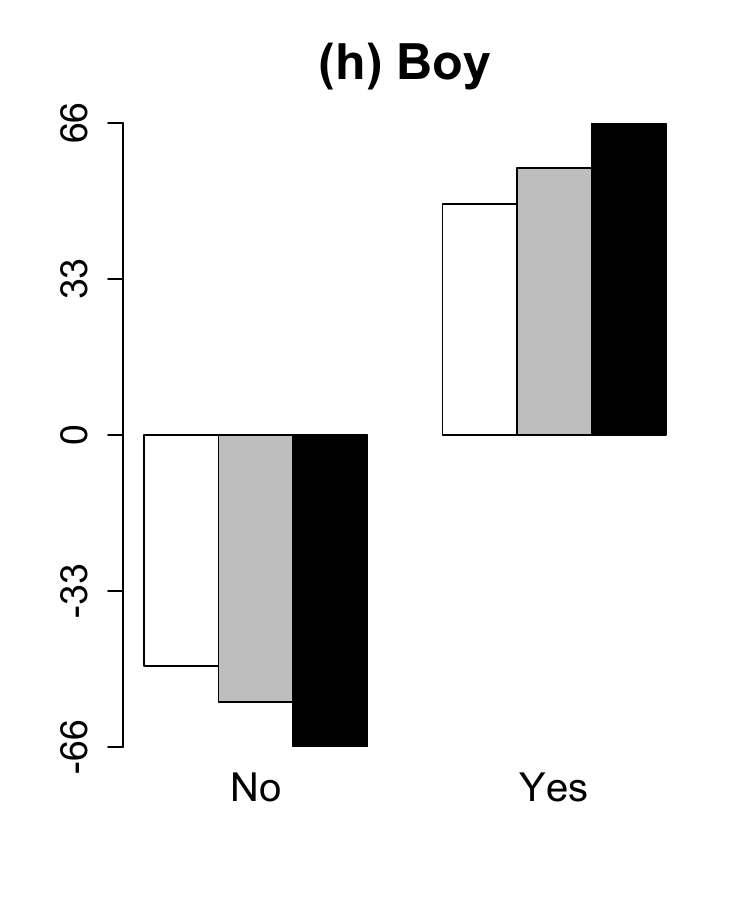}}
\centerline{
\includegraphics[page=9,width=0.25\textwidth]{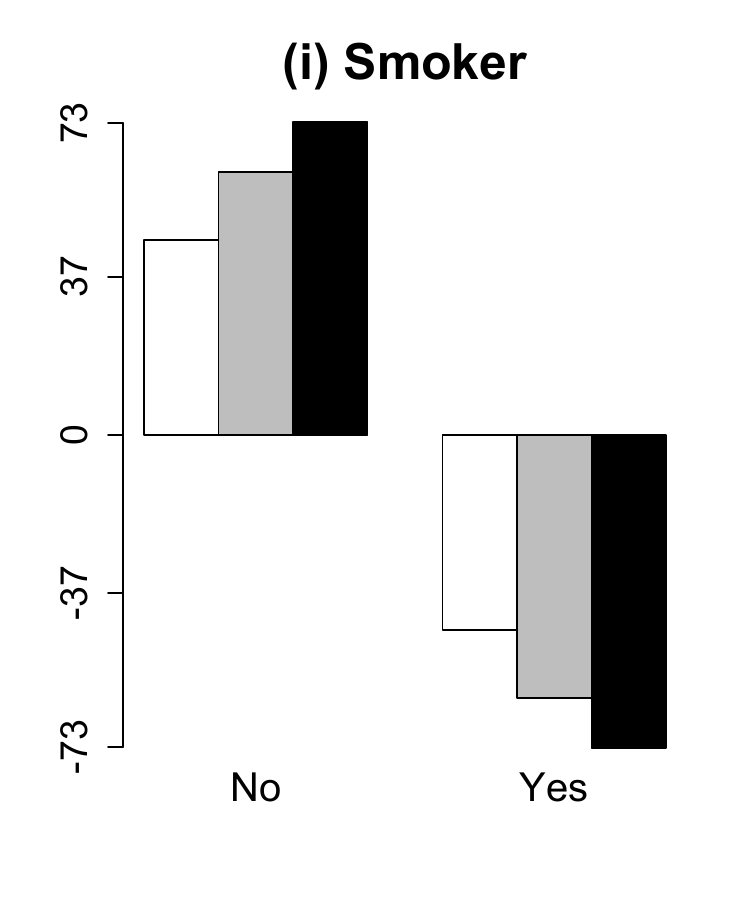}
\includegraphics[page=11,width=0.375\textwidth]{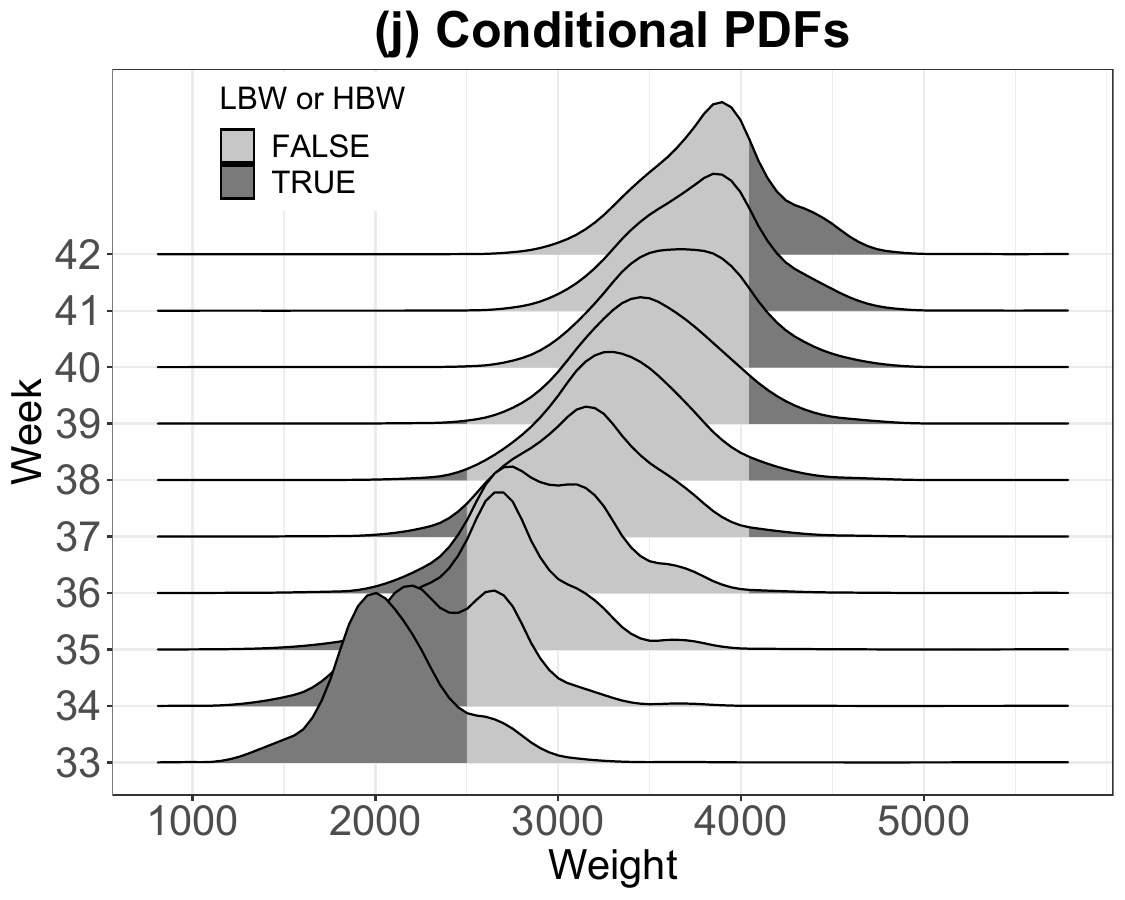}
\includegraphics[page=12,width=0.375\textwidth]{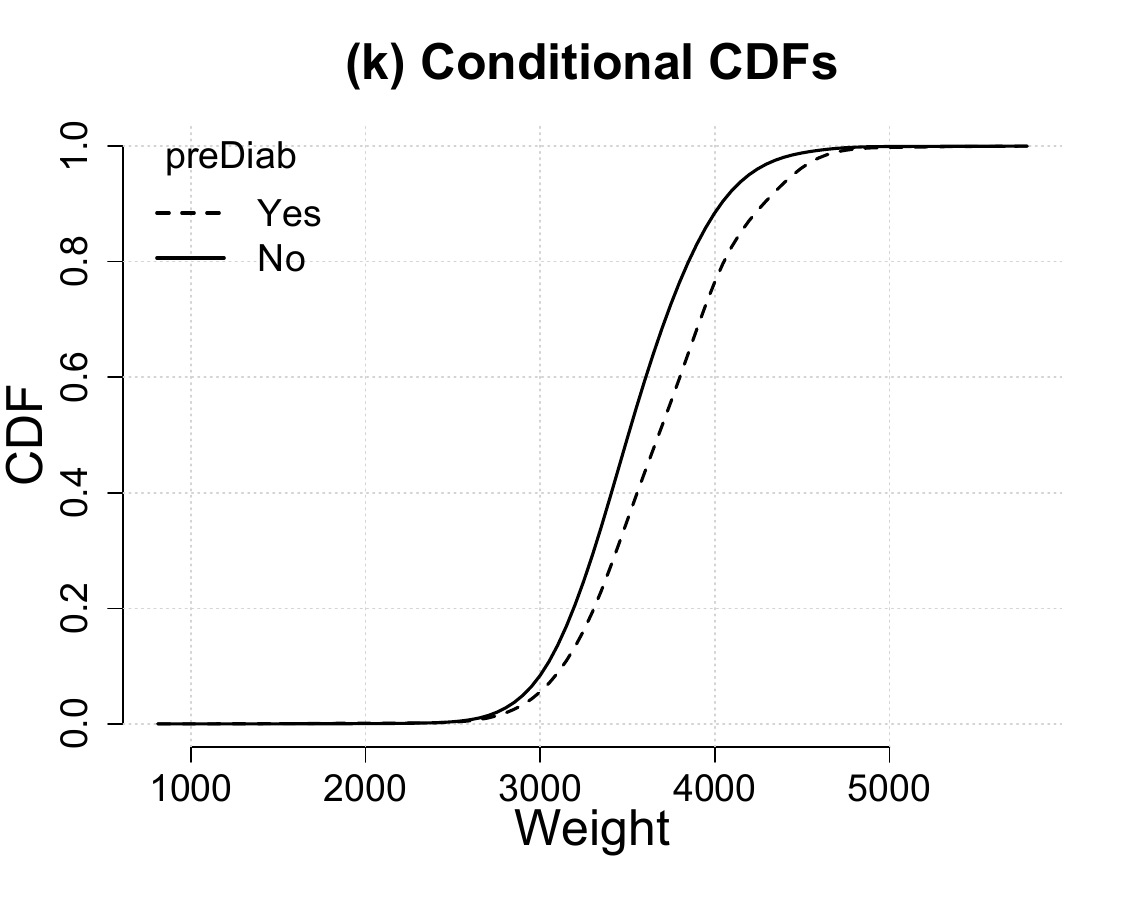}
}
\caption{Marginal main effect analysis of the birth weight data. (a)--(i) Posterior mean ALE main effects at $\tau\in\{0.05,0.50,0.95\}$ for the top 9 important covariates. For continuous covariates, black dashed line represents the value 0. (j)-(i) Conditional distribution estimates by  gestational age (\texttt{Week}) and pre-pregnancy diabetes indicator (\texttt{preDiab}), respectively, with all other covariates fixed at their median (continuous covariates) or mode (binary covariates). }\label{fig:birthwt_main_ale}
\end{figure}

To understand the functional relationship between the top covariates and the predicted birth weight quantiles of QUINN, we plot their estimated ALE main effects $\hat{\bar{Q}}_j(\tau)$ at $\tau\in\{0.05,0.50,0.95\}$ in Figure~\ref{fig:birthwt_main_ale}(a)--(i). The results show that higher values of \texttt{Week}, \texttt{height}, \texttt{BMI}, \texttt{wtGain}, \texttt{preDiab} and \texttt{Boy} are associated with higher predicted birth weight, whereas higher values of \texttt{Cigarette}, \texttt{Black} and \texttt{Smoker} are associated with lower predicted birth weight. Furthermore, the effects of \texttt{Week}, \texttt{height}, \texttt{BMI}, \texttt{wtGain}, \texttt{Cigarette} and \texttt{preDiab} on birth weight are significantly non-constant across quantiles, and the effects of \texttt{Week}, \texttt{BMI}, \texttt{wtGain}, \texttt{Cigarette} are highly non-linear. For example, as \texttt{Cigarette} increases, LBW and TBW display a consistent downward trend, whereas HBW plateaus when \texttt{Cigarette} is greater than 13; as \texttt{wtGain} increases, HBW and TBW display a consistent upward trend, whereas LBW plateaus when \texttt{wtGain} is greater than 65. 

The results in Figure~\ref{fig:birthwt_main_ale}(a)--(i) also provide numerical quantification of the main effects on the predicted birth weight quantiles. For example, Figure~\ref{fig:birthwt_main_ale}(b) shows that compared to mothers who do not smoke during pregnancy, mothers who smoke as many as 20 cigarettes daily are associated with a more than 250-gram decrease in predicted LBW, on average. Figure~\ref{fig:birthwt_main_ale}(g) shows that compared to mothers who do not have pre-existing diabetes, mothers who have pre-existing diabetes are associated with a 254-gram increase in predicted HBW, on average. 

In addition to covariate effects on specific quantiles, QUINN also allows direct characterization of covariate effects on the whole conditional distribution thanks to its density regression nature. To illustrate this property of QUINN,  Figure~\ref{fig:birthwt_main_ale}(j) plots the predicted birth weight density for \texttt{Week} $\in\{33,34,...,42\}$. The result shows that gestational age has a prominent effect on the location of the predicted density. The shifting of the density is most significant when gestational age increases from 33 to 37 and gradually plateaus when the pregnancy term further increases. For each predicted density, we also highlight the region of LBW (\texttt{Weight}$<$2500) and HBW (\texttt{Weight}$>$4000). The result indicates that preterm (\texttt{Week}$<$37) and postterm (\texttt{Week}$>$40) are determinant factors of LBW and HBW, repsectively. Figure~\ref{fig:birthwt_main_ale}(k) plots the predicted birth weight CDF for different levels of \texttt{preDiab}. Compared to mothers who do not have pre-existing diabetes, mothers with pre-existing diabetes are associated with a 10\% increase in probability of giving birth to an overweight infant. 

\begin{figure}
\centerline{\includegraphics[page=13,width=\linewidth]{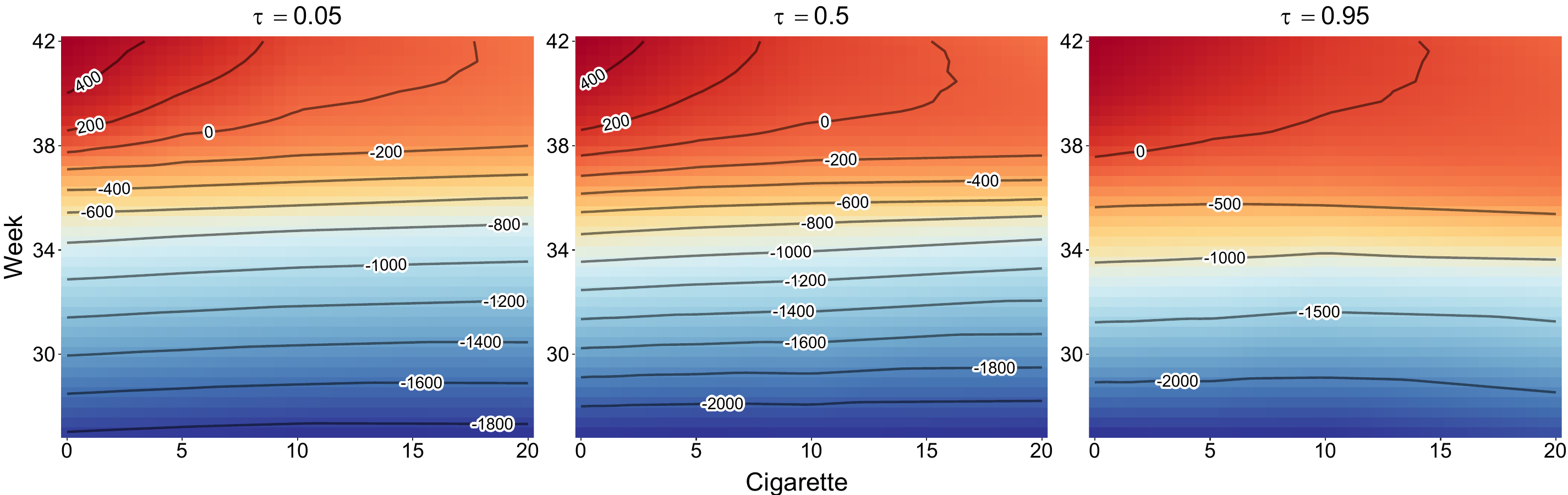}}
\caption{Posterior mean ALE joint effects of gestational age (\texttt{Week}) and average daily number of cigarettes (\texttt{Cigarette}) at $\tau\in\{0.05, 0.5, 0.95\}$. This ﬁgure appears in color in the electronic version of this article.}\label{f:birthwt_int}
\end{figure}

We further analyze the second-order interaction effects between the top covariates that have significant main effects. For each combination, we estimate its ALE joint effect $\bar{Q}_{jl}(\tau)$ as well as its pure interaction effect $\bar{Q}^I_{j,l}(\tau)$ at $\tau\in\{0.05,0.5,0.95\}$. The significance of each interaction is determined by the estimated VI measure $\widehat{\mbox{VI}}_{jl}(\tau)$. Among all interaction considered, the interaction between gestational age and average daily number of cigarettes during pregnancy (\texttt{Week} $\times$ \texttt{Cigarette}) yields the highest VI measure on all three quantiles. To understand the functional relationship between \texttt{Week} $\times$ \texttt{Cigarette} and the birth weight quantiles, Figure~\ref{f:birthwt_int} plots the contour of the joint effects of \texttt{Week} and \texttt{Cigarette}, as quantified by $\bar{Q}_{jl}(\tau)$. The result indicates that higher gestational age is associated with higher birth weight irregardless of maternal smoking habit, but the effect is clearly amplified for non-smokers. In addition, heavier maternal smoking is associated with lower birth weight only for births that occur after the 37th week of pregnancy.

\section{Conclusion}
\label{s:conc}
In this paper, we propose a novel non-linear SQR model that leverges the flexibility of spline and neural network. We adopt a Bayesian framework by assigning prior distributions to the weight parameters and utilize the state-of-art NUTS to sample efficiently from the high-dimensional posterior. Compared to existing 
works, our method models the full quantile process, does not involve constrained optimization, and scales to high-dimensional setting. We also show that our model can yield meaningful interpretation via ALE plots and variable importance scores. Simulation studies show that our model better recover high-dimensional quantile process with complex structure and is robust to data and model sparsity. Sensitivity analysis shows that our can accurately captures quantile-dependent covariate effects.

%The proposed model was applied to the intensity data of tropical cyclones in the North Atlantic Basin and specifically to investigate how intensity quantiles changed between years 1981 and 2006. Our results showed that cyclone intensity increased smoothly from 1981 to 2006, with intensity of the strongest cyclones increasing more significantly than that of typical and weak cyclones. 
The proposed method was used to analyze the relationship between birth weight and pregnancy-related factors of U.S. newborns and specifically to identify influential effects on LBW and HBW. Our results showed that LBW is primarily associated with prematurity and heavy maternal smoking; whereas HBW is primarily associated high maternal body mass index, maternal height, maternal weight gain, and having pre-existing diabetes. 
Future extension can focus on accommodating spatial and/or temporal correlation between observations and variable/model selection using sparsity-inducing priors.

\begin{singlespace}
	\bibliographystyle{rss}
	\bibliography{refs}
\end{singlespace}

\appendix
\renewcommand{\thesection}{\Alph{section}}
\setcounter{section}{0}

\section*{Appendices}

\section{MCMC sampling details}\label{app:mcmc}

\subsection{Posterior evaluation}
The non-parametric model on the conditional PDF of $Z$ given $\X=\bx$ is
\begin{equation*}
\begin{split}
        f_Z(z|\bx,\mathcal{W}) &= \sum_{m=1}^{r+p-1}\theta_m(\bx,\mathcal{W})M_{m,r}(z|\bT) =  \sum_{m=1}^{r+p-1}\frac{\exp\{u_m(\bx,\mathcal{W})\}}{\sum_{i=1}^{r+p-1}\exp\{u_i(\bx,\mathcal{W})\}}M_{m,r}(z|\bT),
\end{split}
\end{equation*}
and the likelihood function is
\begin{equation*}
    \mathcal{L}(\calD|\mathcal{W})=\prod_{i=1}^nf_Z(z_i,\bx_i|\mathcal{W}) = \prod_{i=1}^n \lcbk \sum_{m=1}^{r+p-1}\frac{\exp\{u_m(\bx_i,\mathcal{W})\}}{\sum_{j=1}^{r+p-1}\exp\{u_j(\bx_i,\mathcal{W})\}}M_{m,r}(z_i|\bT)\rcbk.
\end{equation*}
where $\calD=\{z_i,\bx_i\}_{i=1}^n$ denotes the observed data. Let $\bm{\Theta}=\{\mathcal{W},\sigma_w,\gamma\}$ denote the set of modeling parameters and hyper-parameters, then the posterior of QUINN is
\begin{equation*}
    \begin{split}
         f(\bm{\Theta}|\calD)\propto\mathcal{N}^+(\gamma|0,a^2)&\prod_{m=1}^{r+p-1}\prod_{l=0}^{V}\mathcal{N}(W_{2ml}|0,\gamma^2)\prod_{j=0}^{d}\mathcal{N}^+(\sigma_j|0,a^2)\\
         &\prod_{l=0}^V\prod_{j=0}^{d}\mathcal{N}(W_{1lj}|0,\sigma_j^2)\prod_{i=1}^nf_Z(z_i,\bx_i|\mathcal{W})
    \end{split}
\end{equation*}
which can be approximated using MCMC methods. Sampling from this posterior is challenging for traditional MCMC methods such as random-walk Metropolis \citep{metropolis1953} and Gibbs sampler \citep{geman1984}. These methods, although straightforward to implement, do not scale well to complicated posterior with high-dimensional parameter space. The former explores the posterior via inefficient random walks, resulting in low acceptance rate and wasted samples; the latter requires knowing the conditional distribution of each parameter, which can be unrealistic in high-dimensional case.

\subsection{Hamiltonian Monte Carlo}\label{ss:hmc}
Hamiltonian Monte Carlo (HMC) \citep{neal2011,betancourt2015,betancourt2017} is a variant of MCMC that permits efficient sampling from a high-dimensional target distribution, provided that all model parameters are continuous. It has gained increasing popularity for its recent applications in inference of Bayesian neural networks \citep{neal2012}. By introducing auxillary variables $\bm{r}$, HMC transforms the problem of sampling from $f(\bm{\Theta}|\calD)$ to sampling from the joint distribution $f(\bm{r},\bm{\Theta}|\calD)= f(\bm{r}|\bm{\Theta},\calD)f(\bm{\Theta}|\calD)$ where $f(\bm{r}|\bm{\Theta},\calD)$ is the auxiliary distribution often assumed to be multivariate Normal and independent of $\bm{\Theta}$ and $\calD$, i,e. $f(\bm{r}|\bm{\Theta},\calD)=f(\bm{r})$. The joint distribution defines a Hamiltonian
\begin{equation*}
\begin{split}
    H(\bm{r},\bm{\Theta}|\calD) &= T(\bm{r})+V(\bm{\Theta}|\calD)\\
    T(\bm{r}) &\coloneqq -\log f(\bm{r})\\
    V(\bm{\Theta}|\calD) &\coloneqq -\log f(\bm{\Theta}|\calD).\label{e:9}
\end{split}
\end{equation*}
which can be used to generate states, i.e. samples of $\bm{\Theta}$ and $\bm{r}$, by simulating the Hamiltonian dynamics
\begin{equation*}
    \frac{\partial \bm{\Theta}}{\partial t}=\frac{\partial T}{\partial \bm{r}}, \ \frac{\partial \bm{r}}{\partial t}=\frac{\partial V}{\partial \bm{\Theta}}\label{e:10}.
\end{equation*}
At any state $(\bm{\Theta}_t, \bm{r}_t)$, HMC proposes the next state $(\bm{\Theta}_{t+L\Delta t}, \bm{r}_{t+L\Delta t})$ by simulating Hamiltonian dynamics for time $L\Delta t$, which is approximated by applying the leapfrog algorithm $L$ times each with step size $\Delta t$. Starting from an initial state, this process is repeated and the visited states form a Markov chain. Compared to random-walk Metropolis, HMC explores the target distribution more efficiently by using gradient of the log-posterior to direct each transition of the Markov chain. Although each step is more computationally expensive than a Metropolis proposal, the Markov chain produced by HMC often yields more distant samples and significantly higher acceptance rate. Although HMC has a high potential, its practical performance depends highly on the values of $L$ and $\Delta t$. Poor choice of either parameter will result in unsatisfactory exploration of the posterior. In this paper, instead of using the original HMC which only allows manual setting of $L$ and $\Delta t$, we use the No-U-Turn Sampler (NUTS). NUTS is an extension to HMC that implements automatic tuning of $L$. Furthermore, we use the dual averaging algorithm to adaptively select $\Delta t$. A detailed description of NUTS with dual averaging is presented in Algorithm 6 of \citet{hoffman2014}.

NUTS is implemented in many probabilistic programming framework, such as PyMC3 \citep{salvatier2016} and Stan \citep{stan2019}. Computational complexity of an HMC implementation is contingent on gradient calculation of the log-posterior, which the aforementioned high-level frameworks handle via automatic differentiation. In our experiment, we observe that automatic differentiation can be extremely time consuming for a posterior as high-dimensional as ours. As a result, we use a low-level R implementation \citep{monnahan2018} that accepts analytic gradients which we manually calculate.  

\subsection{Reparametrization and transformation}
The hierarchical Gaussian priors $W_{1vw}\indep\mathcal{N}(0,\sigma_w^2)$, $\sigma_w\iid\mathcal{N^+}(0,a^2)$ introduce strong correlation between $W_{1vw}$ and $\sigma_w$ in the posterior, especially when the data size is small. To alleviate this issue, we consider a reparametrization:
\begin{equation*}
     B_{1vw}\iid\mathcal{N}(0,1),\
     \sigma_w\iid\mathcal{N^+}(0,a^2),\
     W_{1vw}=\sigma_wB_{1vw}
\end{equation*}
where $B_{1vw}$ can be considered as standardized weights. Because $B_{1vw}$ and $\sigma_w$ follow independent prior distributions, they are marginally uncorrelated in the posterior. Their coupling is instead introduced in the likelihood function. Such a parameterization is called non-centered. Non-centered parameterization leads to simpler posterior geometries, thus increasing the efficiency of HMC. Similarly, $W_{2vw}\indep\mathcal{N}(0,\gamma^2)$, $\gamma\sim\mathcal{N^+}(0,a^2)$ can be reparameterized as
\begin{equation*}
    B_{2vw}\iid\mathcal{N}(0,1),\
    \gamma\sim\mathcal{N^+}(0,a^2),\
    W_{2vw}=\gamma B_{2vw}.
\end{equation*}

HMC requires $\bm{\Theta}$ to lie in an unconstrained space, and thus every parameter that has a natural constraint needs to be transformed to an unconstrained variable. After unconstrained posterior samples are drawn, they can be back-transformed to the constrained space. In $\bm{\Theta}$, the scale parameters $\sigma_v$ and $\gamma$ are naturally constrained to be positive. Therefore we work with their log-transformations $\tilde{\sigma}_v=\log\sigma_v$ and $\tilde{\gamma}=\log\gamma$ with transformed prior distributions
\begin{equation*}
    f(\tilde{\sigma}_v)=\mathcal{N}^+(\exp(\tilde{\sigma}_v)|0,a^2)\exp(\tilde{\sigma}_v)\mbox{  \ \ \ \ and \ \ \ \ }f(\tilde{\gamma})=\mathcal{N}^+(\exp(\tilde{\gamma})|0,a^2)\exp(\tilde{\gamma}).
\end{equation*}

Let $\mathcal{B}=\{\beta_{uvw}\}$ and $\tilde{\bm{\Theta}}=\{\mathcal{B},\tilde{\sigma}_w,\tilde{\gamma}\}$, then the posterior after non-centered reparameterization and constraint transformation is
\begin{equation*}
    \begin{split}
        f(\tilde{\bm{\Theta}}|\calD)\propto\mathcal{N}^+(\exp(\tilde{\gamma})|0,a^2)\exp(\tilde{\gamma})&\prod_{m=1}^{r+p-1}\prod_{l=0}^{V}\mathcal{N}(B_{2ml}|0,1)\prod_{j=0}^{d}\mathcal{N}^+(\exp(\tilde{\sigma})_j|0,a^2)\exp(\tilde{\sigma}_j)\\
         &\prod_{l=0}^V\prod_{j=0}^{d}\mathcal{N}(B_{1lj}|0,1)\prod_{i=1}^nf_Z(z_i,\bx_i|\mathcal{W}),
    \end{split}
    \label{e:pos}
\end{equation*}
where $W_{1vw}=\exp(\tilde{\sigma}_w)B_{1vw}$ and $W_{2vw}=\exp(\tilde{\gamma})B_{2vw}$ in the likelihood function.
\subsection{Analytic gradient}
In this section, we provide analytic formulas for computing the gradient of $\log f(\tilde{\bm{\Theta}}|\calD)$, which NUTS uses to generate samples of $\tilde{\bm{\Theta}}$. To start with, let $\bX$ denote the observed covariate matrix, $\bM_r(\bz|\bT)$ denote the M-spline matrix of transformed response vector $\bz$, $\bm{1}$ denote a column vector of 1s, $\bm{\sigma}$ denote the vector with elements $\sigma_w$, and $\bB_u$ denote the matrix with elements $B_{uvw}$. The log-likelihood function parametrized by $\tilde{\bm{\Theta}}$ can be written in a compact form using matrix notation
\begin{equation*}
    \begin{split}
       \ell(\calD|\tilde{\bm{\Theta}})&=\sum_{i=1}^n\lpth \log\lsbk \sum_{m=1}^{r+p-1}\exp \lcbk u_m(\bx_i,\bm{\tilde{\Theta}})\rcbk M_{m,r}(z_i|\bT)\rsbk-\log\lsbk \sum_{m=1}^{r+p-1}\exp \lcbk u_m(\bx_i,\tilde{\bm{\Theta}})\rcbk\rsbk\rpth\\\\
       &=\bm{1}^T\lpth \log\lsbk \exp\lcbk\bU(\bX,\bm{\tilde{\Theta}})\rcbk\odot\bM_r(\bz|\bT)\bm{1}\rsbk-\log\lsbk \exp\lcbk\bU(\bX,\bm{\tilde{\Theta}})\rcbk\bm{1}\rsbk\rpth
    \end{split}
\end{equation*}
where
\begin{equation*}
    \bU(\bX,\bm{\tilde{\Theta}})=\begin{amatrix}{1}
        \bm{1} & \phi\lcbk\tilde{\bX}\diag[\exp(\tilde{\bm{\sigma}})]\bB_1\rcbk
        \end{amatrix}\lsbk\exp(\tilde{\gamma})\bB_2\rsbk,\
    \tilde{\bX}=\begin{amatrix}{1}
        \bm{1} & \bX\end{amatrix},
\end{equation*}
and $\diag[\exp(\tilde{\bm{\sigma}})]$ is the diagonal matrix with diagonal entries $\exp(\tilde{\bm{\sigma}})$. The log-prior can be written as
\begin{equation*}
    f(\tilde{\bm{\Theta}})\propto-\frac{\exp(2\tilde{\gamma})a^2}{\pi}+\tilde{\gamma}-\frac{\vect(\bB_2)^T\vect(\bB_2)}{2}-\bm{1}^T\lsbk\frac{\exp(2\tilde{\bm{\sigma}})a^2}{\pi}-\tilde{\bm{\sigma}}\rsbk-\frac{\vect(\bB_1)^T\vect(\bB_1)}{2}.
\end{equation*}
where $\vect(\cdot)$ denotes the vectorization operator. Finally, the gradient formula of the log-posteior with respect to each parameter, expressed using matrix notation, is given by
\begin{equation*}
    \begin{split}
    \frac{\partial\log f(\tilde{\bm{\Theta}}|\calD)}{\partial  \bB_1}=&\exp(\tilde{\gamma})\diag\lsbk \exp(\tilde{\bm{\sigma}})\rsbk\tilde{\bX}^T\lpth \bV_1\bV_3-\bV_2\bV_3\rpth-\bB_1\\
    \frac{\partial\log f(\tilde{\bm{\Theta}}|\calD)}{\partial  \bB_2}=&\exp(\tilde{\gamma})\bV_0^T\lpth\bV_1\lsbk\exp\lcbk\bU(\bX,\bm{\tilde{\Theta}})\rcbk\odot\bM_r(\bz|\bT)\rsbk-\bV_2\exp\lcbk\bU(\bX,\bm{\tilde{\Theta}})\rcbk\rpth-\bB_2\\
    \frac{\partial\log f(\tilde{\bm{\Theta}}|\calD)}{\partial\tilde{\bm{\sigma}}}=&\exp(\tilde{\gamma})\lsbk\diag\lcbk \tilde{\bX}^T\lpth\bV_1\bV_3-\bV_2\bV_4\rpth\bB_1^T\rcbk\rsbk\odot\exp(\tilde{\bm{\sigma}})
    -\frac{2a^2}{\pi}\exp(2\tilde{\bm{\sigma}})+\bm{1}\\
    \frac{\partial\log f(\tilde{\bm{\Theta}}|\calD)}{\partial  \tilde{\gamma}}=&\exp(\tilde{\gamma})\Bigg(\tr\lcbk\bV_0\bB_2\lsbk\exp\lcbk\bU(\bX,\bm{\tilde{\Theta}})\rcbk\odot\bM_r(\bz|\bT)\rsbk^T\bV_1\rcbk\\
    &-\tr\lcbk\bV_0\bB_2\exp\lcbk\bU(\bX,\bm{\tilde{\Theta}})\rcbk^T\bV_2\rcbk\Bigg)-
    \frac{2a^2}{\pi}\exp(2\tilde{\gamma})+1
    \end{split}
\end{equation*}
where
\begin{equation*}
\begin{split}
\bV_0 =& \begin{amatrix}{1}\bm{1} & \phi\lcbk\tilde{\bX}\diag[\exp(\tilde{\bm{\sigma}})]\bB_1\rcbk\end{amatrix}\\
\bV_1 =& \diag \lcbk \bm{1}\oslash\lsbk\exp\lcbk\bU(\bX,\bm{\tilde{\Theta}})\rcbk\odot\bM_r(\bz|\bT)\bm{1}\rsbk\rcbk\\
\bV_2 =& \diag \lcbk \bm{1}\oslash\lsbk \exp\lcbk\bU(\bX,\bm{\tilde{\Theta}})\rcbk\bm{1}\rsbk\rcbk\\
\bV_3 =& \lsbk \exp\lcbk\bU(\bX,\bm{\tilde{\Theta}})\rcbk\odot\bM_r(\bz|\bT)\bar{\bB}_2^T\rsbk\odot\phi'\lpth \tilde{\bX}\diag[\exp(\tilde{\bm{\sigma}})]\bB_1\rpth\\
\bV_4 =& \lsbk \exp\lcbk\bU(\bX,\bm{\tilde{\Theta}})\rcbk\bar{\bB}_2^T\rsbk\odot\phi'\lpth \tilde{\bX}\diag[\exp(\tilde{\bm{\sigma}})]\bB_1\rpth,
\end{split}
\end{equation*}
$\odot$ denotes element-wise multiplication, $\oslash$ denotes element-wise division, and $\bar{\bB}_u$ is $\bB_u$ after removing the first row.
\subsection{Model estimation}
It is well-known that FNN suffers from over-parameterization, which makes the weight parameters highly non-identifiable. In practice, MCMC for individual weights might not even converge, making Bayesian inference of the weight parameters impossible. Let $\mathcal{W}^{(t)},t=1,...,T$ denote the $t$-th posterior sample of $\mathcal{W}$. In this study, instead of using the posterior estimates (e.g. posterior means) of the weight parameters to calculate a single estimate of $F_Z(z|\bx,\hat{\mathcal{W}})$
\begin{equation*}
        \hat{\mathcal{W}} = \frac{1}{T}\sum_{t=1}^T\mathcal{W}^{(t)}
\end{equation*}
we estimate $F_Z(z|\bx)$ using its posterior mean
\begin{equation*}
    \hat{F}_Z(z|\bx)=\frac{1}{T}\sum_{t=1}^TF(z|\bx,\mathcal{W}^{(t)}).
\end{equation*}
Convergence of MCMC can be checked using the trace plot of $F_Z(z|\bx,\mathcal{W}^{(t)})$ for some $(z,\bx)$.
\subsection{Convergence diagnostics}
To monitor the convergence of NUTS, we simulate multiple independent chains and inspect the trace plots of data log-likelihood. We choose to monitor the log-likelihood because it is efficient to calculate  and represents the goodness-of-fit of our non-parametric model. We do not monitor individual weight parameters because they suffer from non-identifiability due to the over-parametrization of FNN; it is likely that their trace plots display multimodality rather than converging to a single distribution. As an illustration, Figure~S\ref{f:trace} plots the traces of log-likelihood of four chains for one replicate of Simulation~1--4, after discarding burn-ins. The plot shows a good mixing of the four chains, all converging to the same target distribution.

In addition to trace plots, we also utilize various diagnostic statistics to more precisely assess convergence. \citet{vehtari2021} recently propose to inspect the values of bulk and tail effective sample size (ESS) together with an improved Gelman-Rubin $\hat{R}$ to assess MCMC convergence. Following their recommendation, we consider the chains converge only if both bulk and tail ESS are greater than 100, and $\hat{R}$ is less than 1.05. Calculation of these statistics are carried out by functions implemented in the R package $\texttt{rstan}$.

\section{Computing the sensitivity indices}\label{app:sens}
In this section, we first briefly summarize how the main and interaction ALE can be estimated using sample data; the full description can be seen in \cite{ApleyZhu2020}. We then explain how VI scores can be estimate based on the ALE estimates.

Let $x_{i,j}$ and $\bx_{i,\setminus j}$ denote the $i$th observation of $j$th covariate and all other covariates respectively. The sample range of $X_j$ is partitioned into $K$ intervals $\{N_j(k)=(z_{k-1,j},z_{k,j}]: k=1,2,...,K\}$ where $z_{k,j}$ are chosen as the $k/K$-th sample percentile if $X_j$ is continuous and the unique values otherwise. Then the uncentered effect $\bar{Q}^U_j(\tau,x_j)$ can be estimated by
\begin{equation*}
    \hat{\bar{Q}}^U_j(\tau,x_j) = \sum_{k=1}^{k_j(x_j)}\frac{1}{n_j(k)}\sum_{\{i:x_{i,j}\in N_j(k)\}} \lsbk q_j(\tau,z_{k,j},\bx_{i,\setminus j})-q_j(\tau,z_{k-1,j},\bx_{i,\setminus j})\rsbk,
\end{equation*}
where $k_j(x_j)$ index the interval into which $x_j$ falls, and $n_j(k)$ denotes the number of sample observations $N_j(k)$ contains such that $n=\sum_{k=1}^kn_j(k)$. Finally, $\bar{Q}_{j}(\tau,x_j)$ can be estimated by mean-centering $\hat{\bar{Q}}^U_{j}(\tau,x_j)$, i.e.
\begin{equation*}
    \hat{\bar{Q}}_{j}(\tau,x_j) = \hat{\bar{Q}}^U_{j}(\tau,x_j) - \frac{1}{n}\sum_{k=1}^Kn_j(k)\bar{Q}^U_{j}(\tau,z_{k,j}).
\end{equation*}

For any pair of covariates $\{X_j,X_l\}$, let $\bx_{i,\{j,l\}}$ denote $i$th observation vector of $j$th and $l$th covariate, and $\bx_{i,\setminus \{j,l\}}$ denote all other covariates. The Cartesian product of sample ranges of $X_j$ and $X_l$ can be partitioned into $K^2$ rectangular cells $N_{\{j,l\}}(k,m)=(z_{k-1,j},z_{k,j}]\times(z_{l-1,j},z_{l,j}]$. Then the uncentered effect $\bar{Q}^U_{j,l}(\tau,x_j,x_l)$ can be estimated by
\begin{equation*}
    \begin{split}
    \hat{\bar{Q}}^U_{jl}(\tau,x_j,x_l) =& \sum_{k=1}^{k_j(x_j)}\sum_{m=1}^{k_l(x_l)}\frac{1}{n_{\{j,l\}}(k,m)}\sum_{\{i:\bx_{i,\{j,l\}}\in N_{\{j,l\}}(k,m)\}} \Big[ q_{jl}(\tau,z_{k,j},z_{m,l},\bx_{i,\setminus \{j,l\}})\\
    &-q_{jl}(\tau,z_{k-1,j},z_{m,l},\bx_{i,\setminus \{j,l\}})-\big\{q_{jl}(\tau,z_{k,j},z_{m-1,l},\bx_{i,\setminus \{j,l\}})\\
    &-q_{jl}(\tau,z_{k-1,j},z_{m-1,l},\bx_{i,\setminus \{j,l\}})\big\}\Big],
    \end{split}
\end{equation*}
where $k_j(x_j),k_l(x_l)$ index the cell into which $(x_j,x_l)$ falls, and $n_{\{j,l\}}(k,m)$ denotes the number of sample observations $N_{\{j,l\}}(k,m)$ contains such that $n=\sum_{k=1}^k\sum_{m=1}^Kn_{\{j,l\}}(k,m)$. Similarly, $\bar{Q}_{jl}(\tau,x_j,x_l)$ can be estimated by mean-centering $\hat{\bar{Q}}^U_{j,l}(\tau,x_j,x_l)$, i.e.,
\begin{equation*}
    \hat{\bar{Q}}_{jl}(\tau,x_j,x_l) =\hat{\bar{Q}}^U_{jl}(\tau,x_j,x_l)-\frac{1}{n}\sum_{k=1}^K\sum_{m=1}^Kn_{\{j,l\}}(k,m)\hat{\bar{Q}}^U_{jl}(\tau,z_{k,j},z_{m,l}).
\end{equation*}
Finally, interaction ALE is estimated by $\hat{\bar{Q}}^I_{j,l}(\tau,x_j,x_l)=\hat{\bar{Q}}_{j,l}(\tau,x_j,x_l)-\hat{\bar{Q}}_j(\tau,x_l)-\hat{\bar{Q}}_j(\tau,x_l)$.

Following its definition in Section~3, $\mbox{VI}_j(\tau)$ can be estimated by the sample standard deviation or the sample range of $\hat{\bar{Q}}_j(\tau,z_{k,j})$, i.e.,
\begin{equation*}
\widehat{\mbox{VI}}_j(\tau) = \begin{cases}
\sqrt{\frac{1}{K}\sum_{k=1}^K\lsbk \hat{\bar{Q}}_j(\tau,z_{k,j})-\frac{1}{K}\sum_{k=1}^K\hat{\bar{Q}}_j(\tau,z_{k,j})\rsbk^2} \ \ \ \text{if } X_j \text{ is continuous}\\
 \lcbk \max_k\lsbk \hat{\bar{Q}}_j(\tau,z_{k,j})\rsbk-\min_k\lsbk \hat{\bar{Q}}_j(\tau,z_{k,j})\rsbk\rcbk/4 \ \ \ \text{if } X_j \text{ is categorical}
\end{cases}.
\end{equation*}
By analogy, $\mbox{VI}_{jl}(\tau)$ can be estimated by the sample standard deviation or the sample range of $\hat{\bar{Q}}^I_{jl}(\tau,z_{k,j},z_{m,l})$.

\section{Simulation study implementation details}\label{app:simu}

In this section, we provide implementation details of the simulation study for the competing methods. For MCQRNN, we consider neural networks with a single hidden layer, $V\in\{3,5,8,10,15\}$ hidden neurons, and weight penalty coefficients $\lambda \in \{ e^{-2},e^{-3},...,e^{-6}\}$. We select the best configuration of $V$ and $\lambda$ using 5-fold cross-validation. For NPSQR and NPDFSQR, we follow the guidelines provided by \citet{DasGhoshal2018} and first transform the response variable and covariate(s) into unit intervals using min-max normalization. The response variable and covariate(s) are then expanded using quadratic B-splines with same number of equidistant knots, denoted as $p_{\text{DG}}$. We fit NPSQR with $p_{\text{DG}}\in\{3,4,...,10\}$ and NPDFSQR with $p_{\text{DG}}\in\{5,6,...,10\}$. The optimal $p_{\text{DG}}$ for either model is chosen based on the AIC in which maximum likelihood estimates are replaced by posterior means. SeriesCDE contains four tuning parameters: the number of components of the series expansion in the $Y$ direction $N_Y$, the number of components of the series expansion in the $\X$ direction $N_{\X}$, the bandwidth parameter $\epsilon$ of the Gaussian kernel for constructing the Gram matrix of $\bX$, and the smoothness parameter $\delta$ that controls the bumpiness of the estimated conditional density function. The tuning grids are set as $N_{\X},N_{Y}\in\{1,2,...,n\}$, $\epsilon\in\{e^{-7},e^{-6.5},...,e^{3}\}$, and $\delta\in\{0,0.05,...,0.5\}$. Following the guidelines provided by \citet{IzbickiLee2016}, we first select the best configuration of $N_{\X}$, $N_Y$, and $\epsilon$ using 5-fold cross-validation. We then tune $\delta$ using again 5-fold cross-validation while fixing $N_{\X}$, $N_Y$, and $\epsilon$ at their optimal values.

Since NPDFSQR, and seriesCDE model the distribution function, their results have to be converted to estimates of the quantile function. For NPDFSQR, once we obtain the non-parametric CDF estimate of the transformed response $\hat{F}_Z(z|\bx)$, we evaluate it at 101 equidistant grid-points on the unit interval for each $\bx$. The conditional quantile function $Q_Z(\tau|\bx)$, $\tau\in(0,1)$ can then be estimated by interpolation using $ \lcbk \hat{F}_Z(z_i,\bx)\rcbk_{i=1}^{101}$ as input values and the aforementioned 101 equidistant grid-points as functional output values. Finally, the quantile function of the original response $Q_Y(\tau|\bx)$ can be estimated by reverting the min-max normalization. For seriesCDE, we first convert the non-parametric density function estimate $\hat{f}_Y(y|\bx)$ to $\hat{F}_Y(y|\bx)$ using numerical integration (e.g. trapezoidal rule). We then estimate $Q_Y(\tau|\bx)$ using the aforementioned interpolation approach.

For NPSQR, NPDFSQR, and seriesCDE, the above parameters are used for Simulation~1--4. However, for MCQRNN, we have slightly different tuning parameters for Simulation 4; we consider neural networks with either a single or two hidden layers, $V\in\{3,5,8,10,15\}$ hidden neurons for each hidden layer, and weight penalty coefficients $\lambda \in\{e^{-2}, e^{-3},...,e^{-7}\}$. The best configuration of number of layers, $V$, and $\lambda$ are selected using 5-fold cross-validation. We do not consider neural networks with more than two hidden layers as they are not currently implemented in $\texttt{qrnn}$.

\section{Additional results}
\begin{figure}
\centerline{\includegraphics[page=3,width=0.5\linewidth]{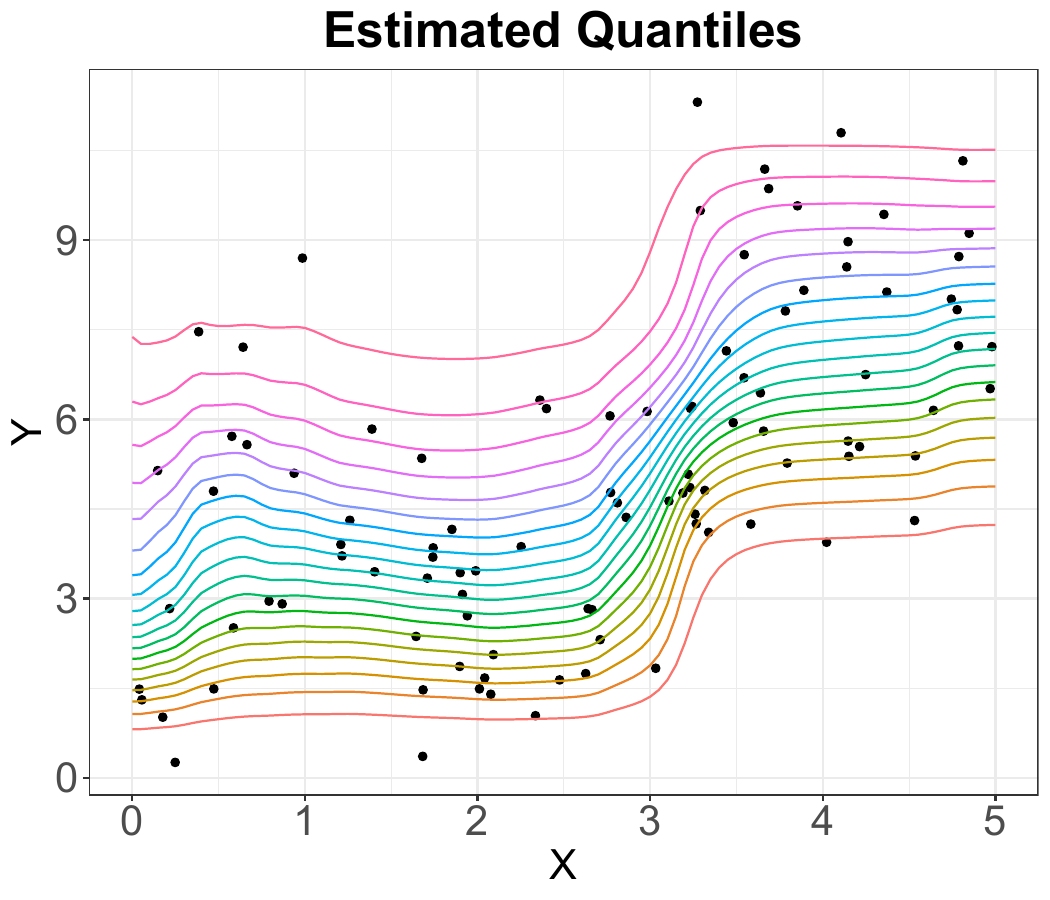}
\includegraphics[page=4,width=0.5\linewidth]{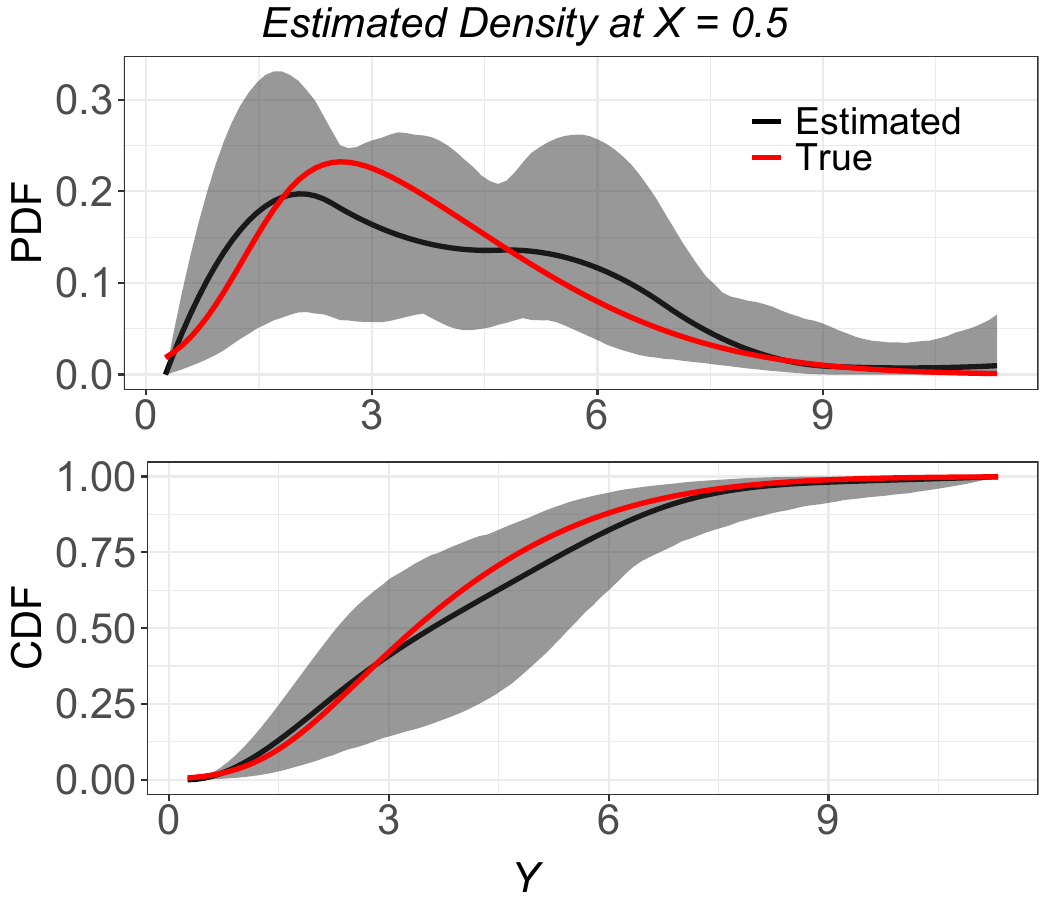}}
\centerline{\includegraphics[page=5,width=0.5\linewidth]{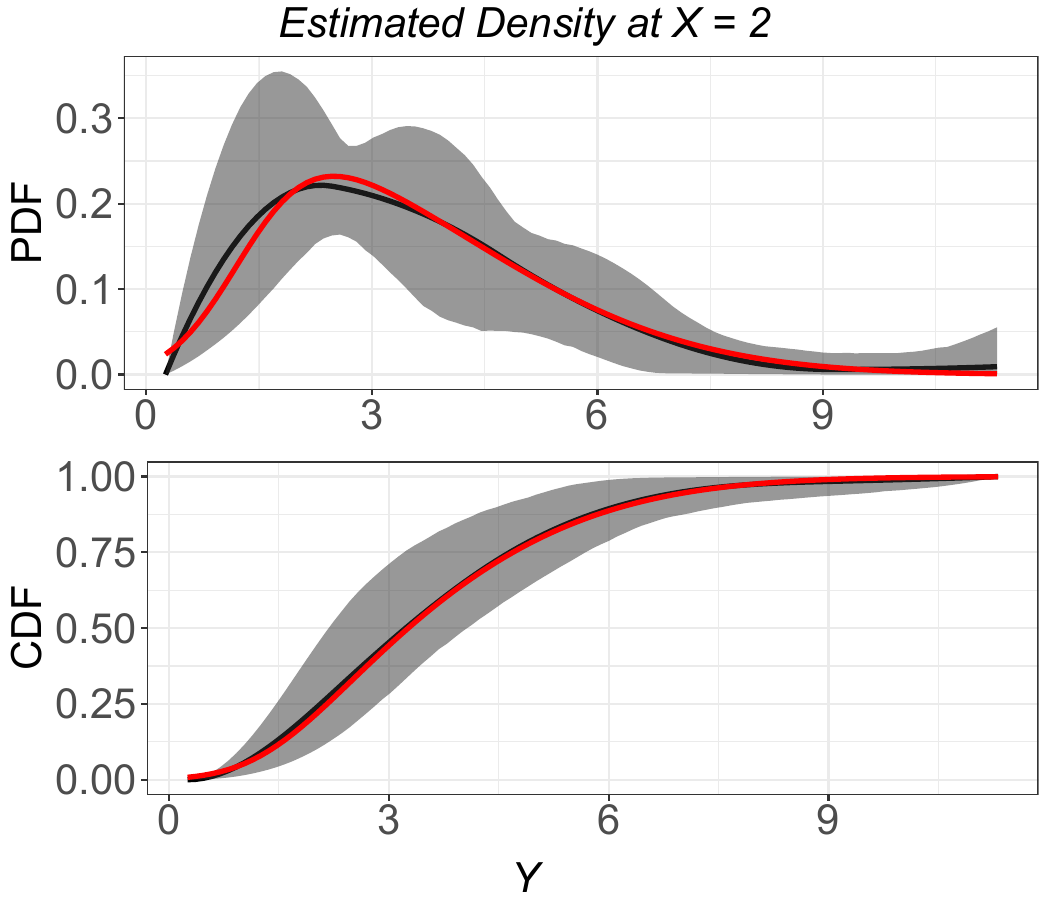}
\includegraphics[page=6,width=0.5\linewidth]{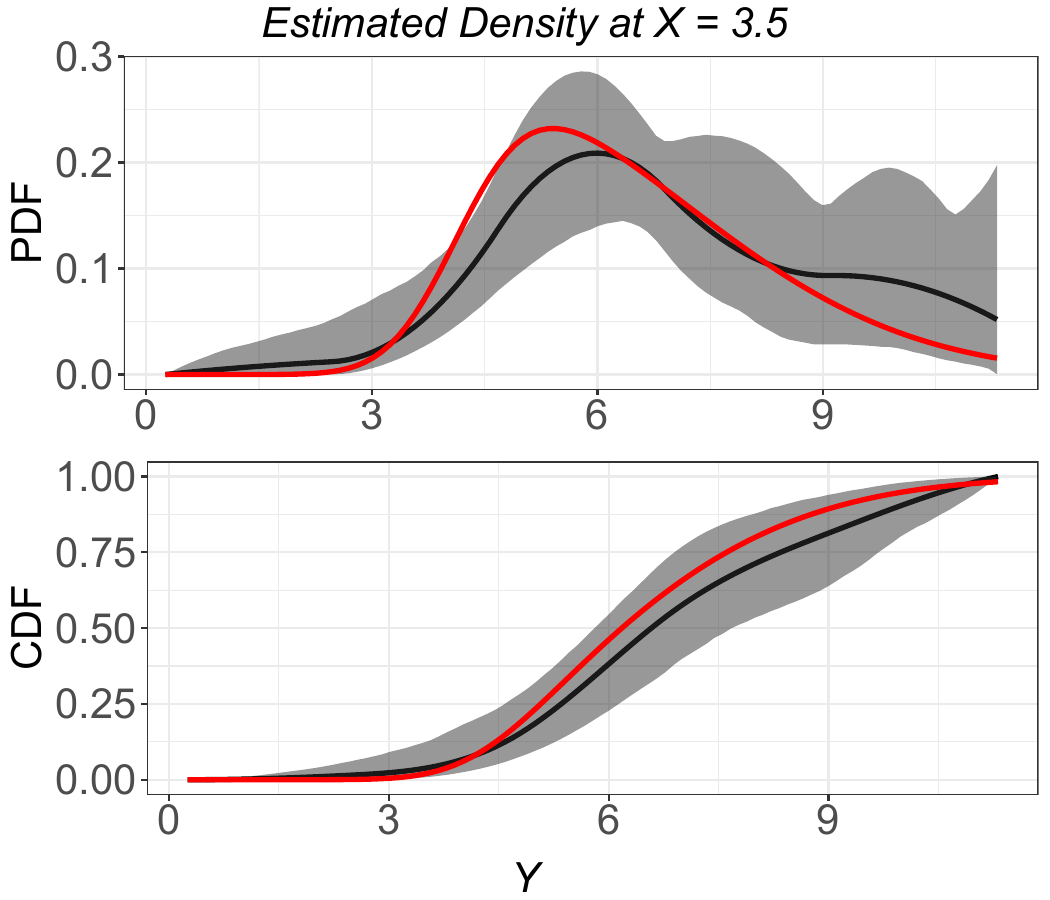}}
\caption{Posterior estimates of quantile curves and conditional density for Simulation 1. Gray shade represents 95\% credible bands. }\label{f:sim1_estimate}
\end{figure}

\begin{figure}
\centerline{\includegraphics[page=7,width=0.5\linewidth]{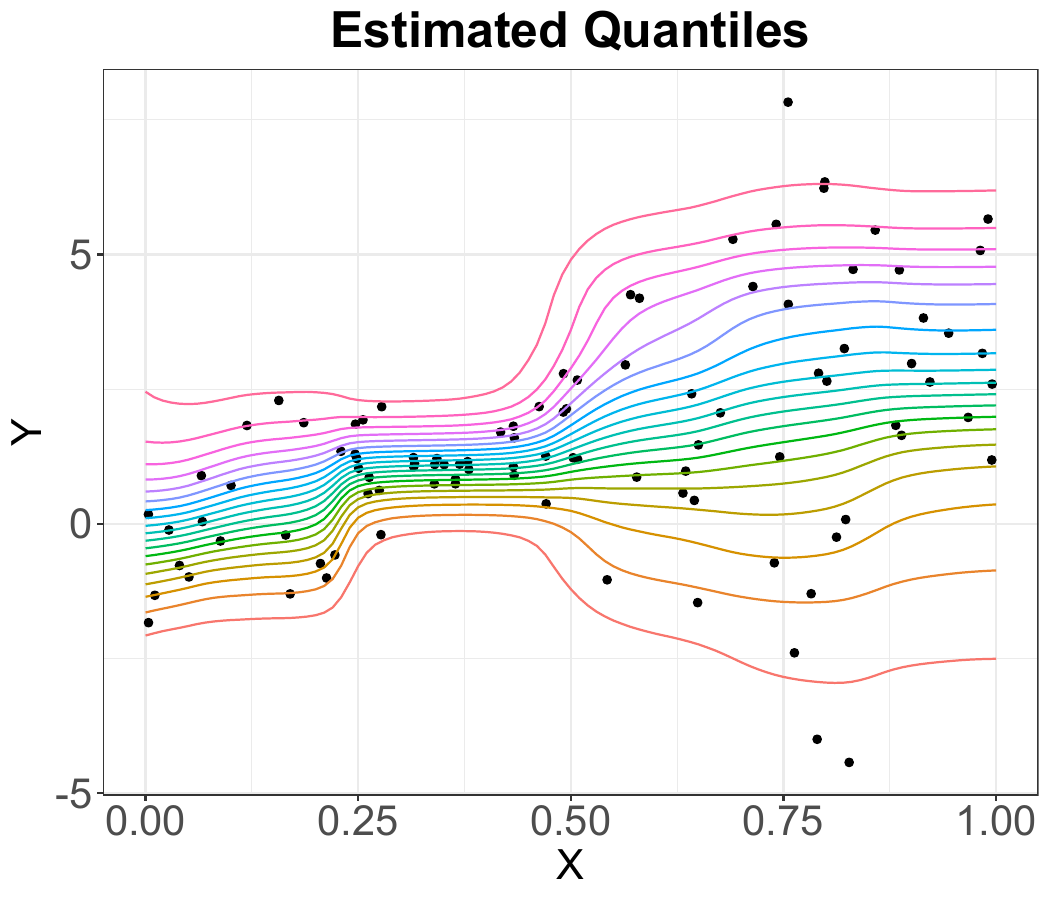}
\includegraphics[page=8,width=0.5\linewidth]{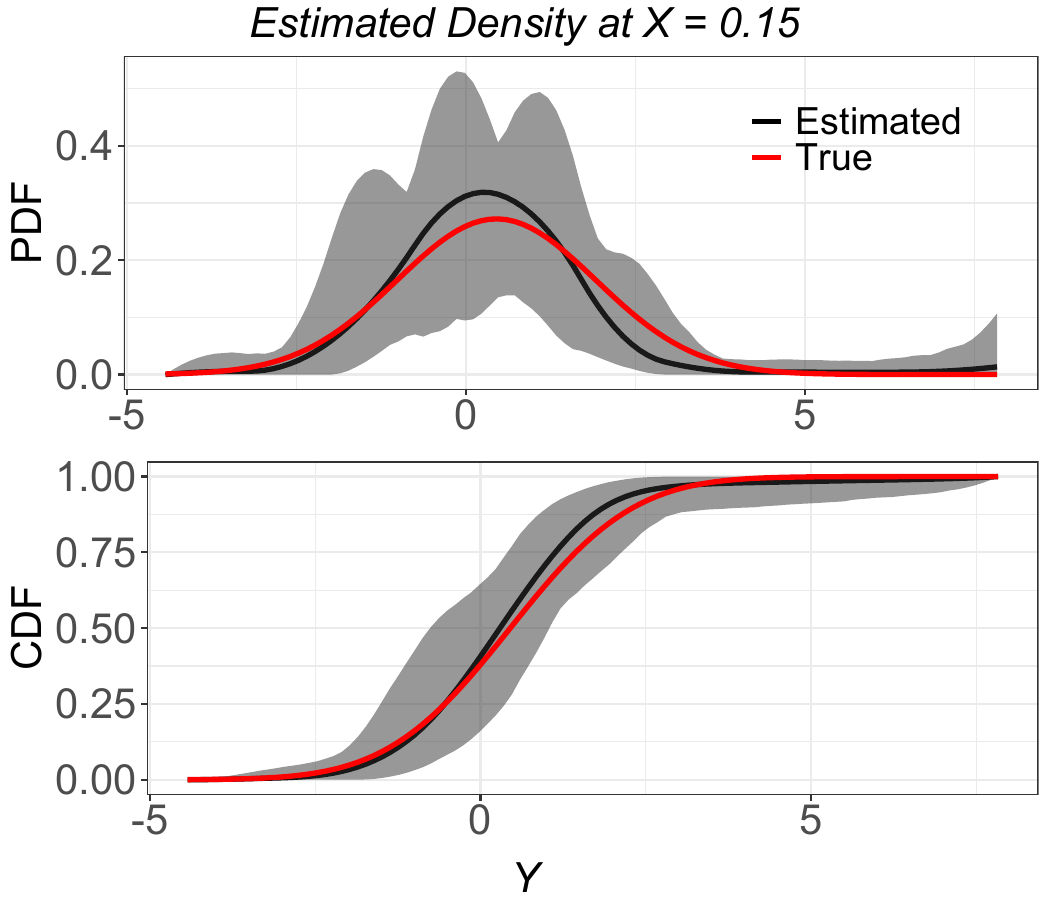}}
\centerline{\includegraphics[page=9,width=0.5\linewidth]{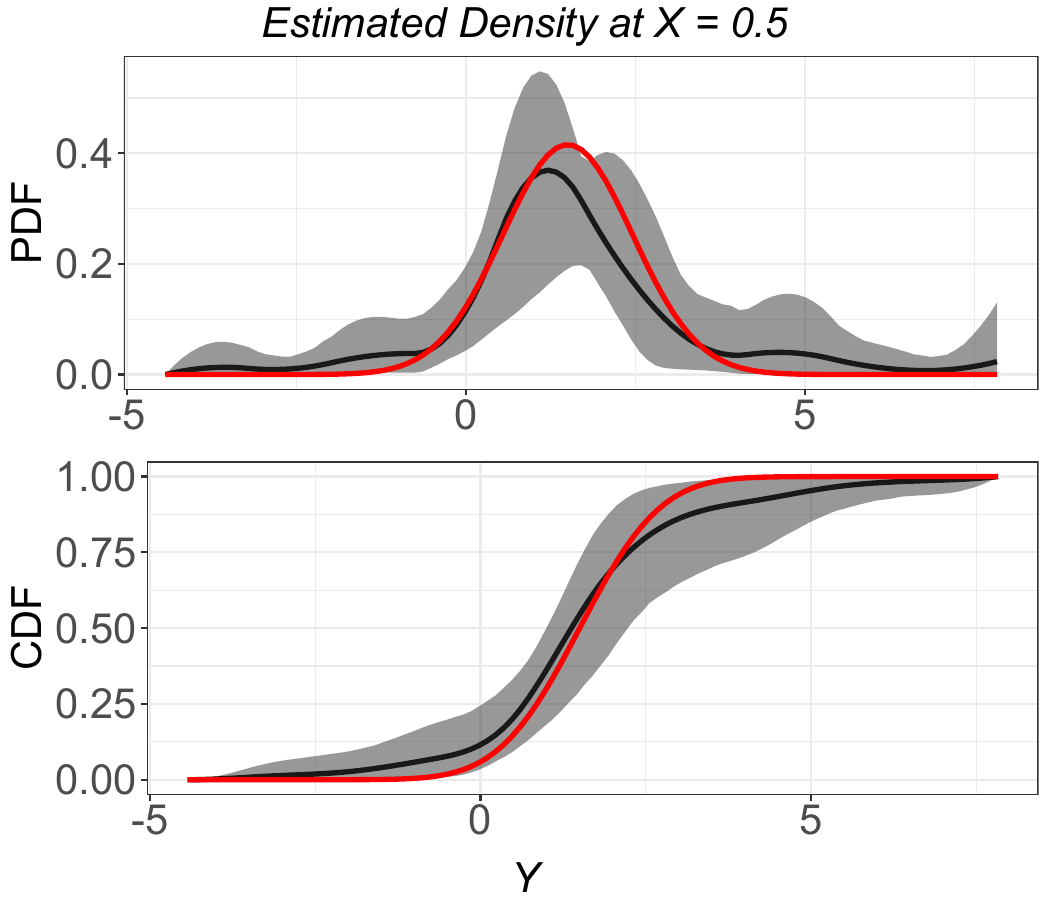}
\includegraphics[page=10,width=0.5\linewidth]{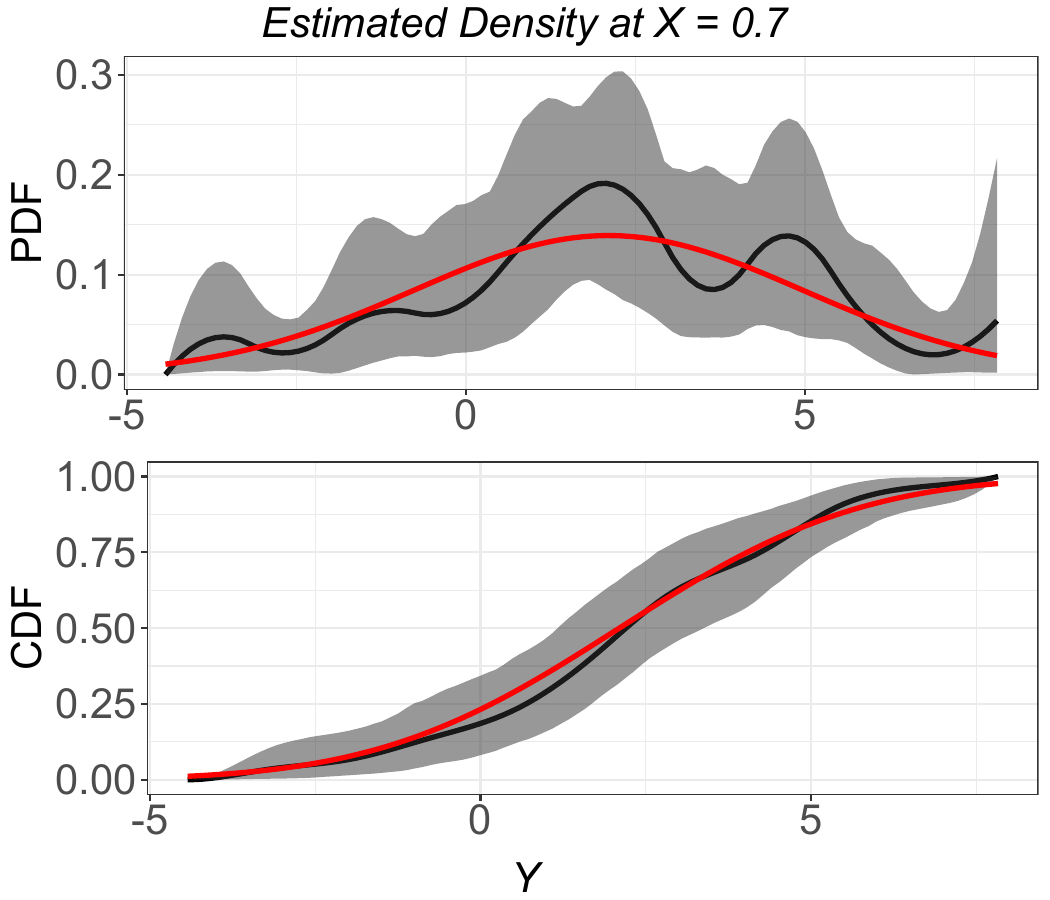}}
\caption{Posterior estimates of quantile curves and conditional density for Simulation 2. Gray shade represents 95\% credible bands.}\label{f:sim2_estimate}
\end{figure}

Figures~\ref{f:sim1_estimate} and \ref{f:sim2_estimate} plot the fitted quantile curves and density functions for one dataset from the first two simulation designs. The remaining figures plot interaction surfaces for eight simulated datasets from the fourth simulation scenario. 

\begin{figure}
\centerline{\includegraphics[page=1,width=\linewidth]{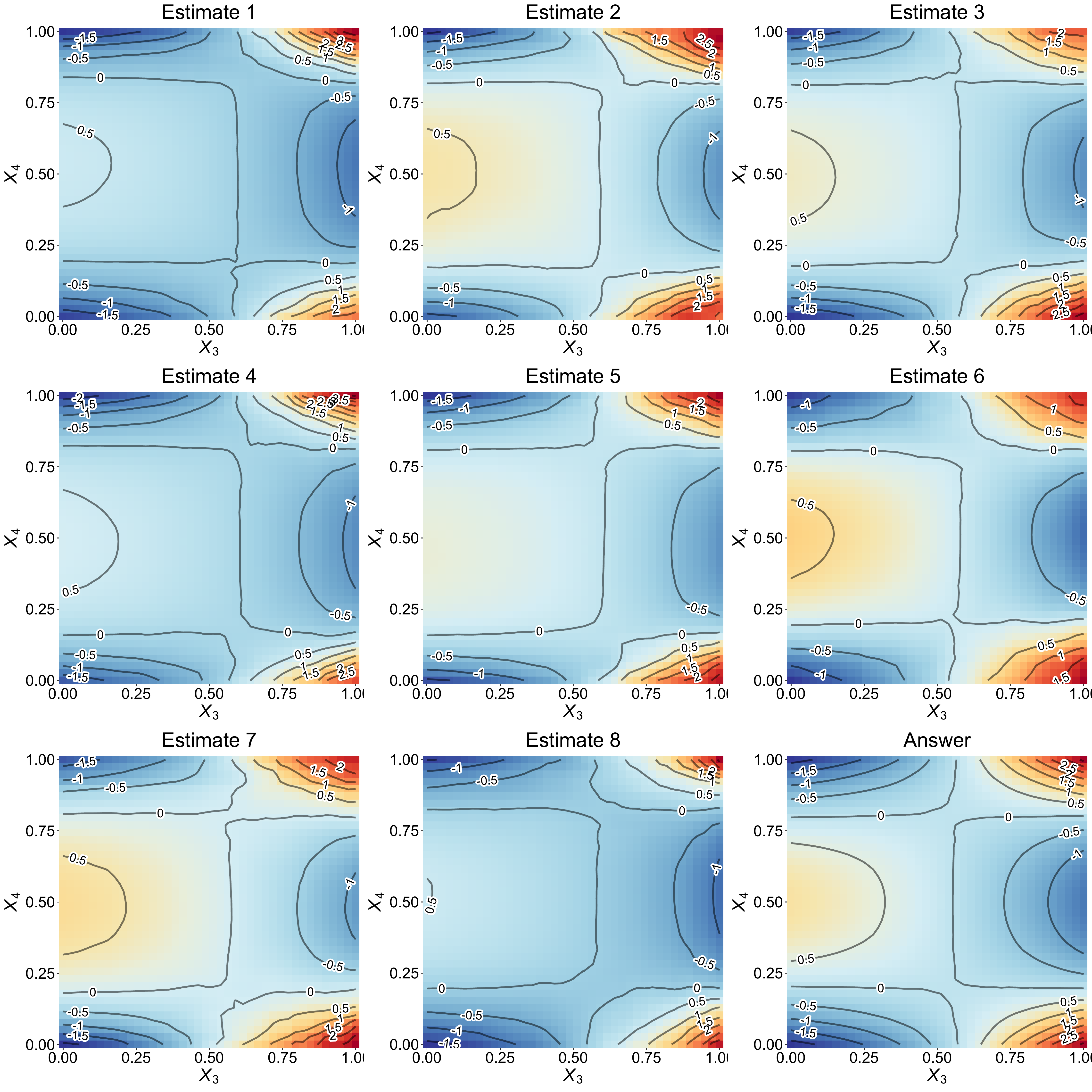}}
\caption{Marginal interaction effect between $X_3$ and $X_4$. Estimate 1--8 are posterior mean ALE interaction effect $\hat{\bar{Q}}^I_{34}(\tau,x_3,x_4)$ of 8 replicates at quantile level $\tau=0.05$. ``Answer" represents the ground truth.}\label{f:sens_int34_1}
\end{figure}

\begin{figure}
\centerline{\includegraphics[page=2,width=\linewidth]{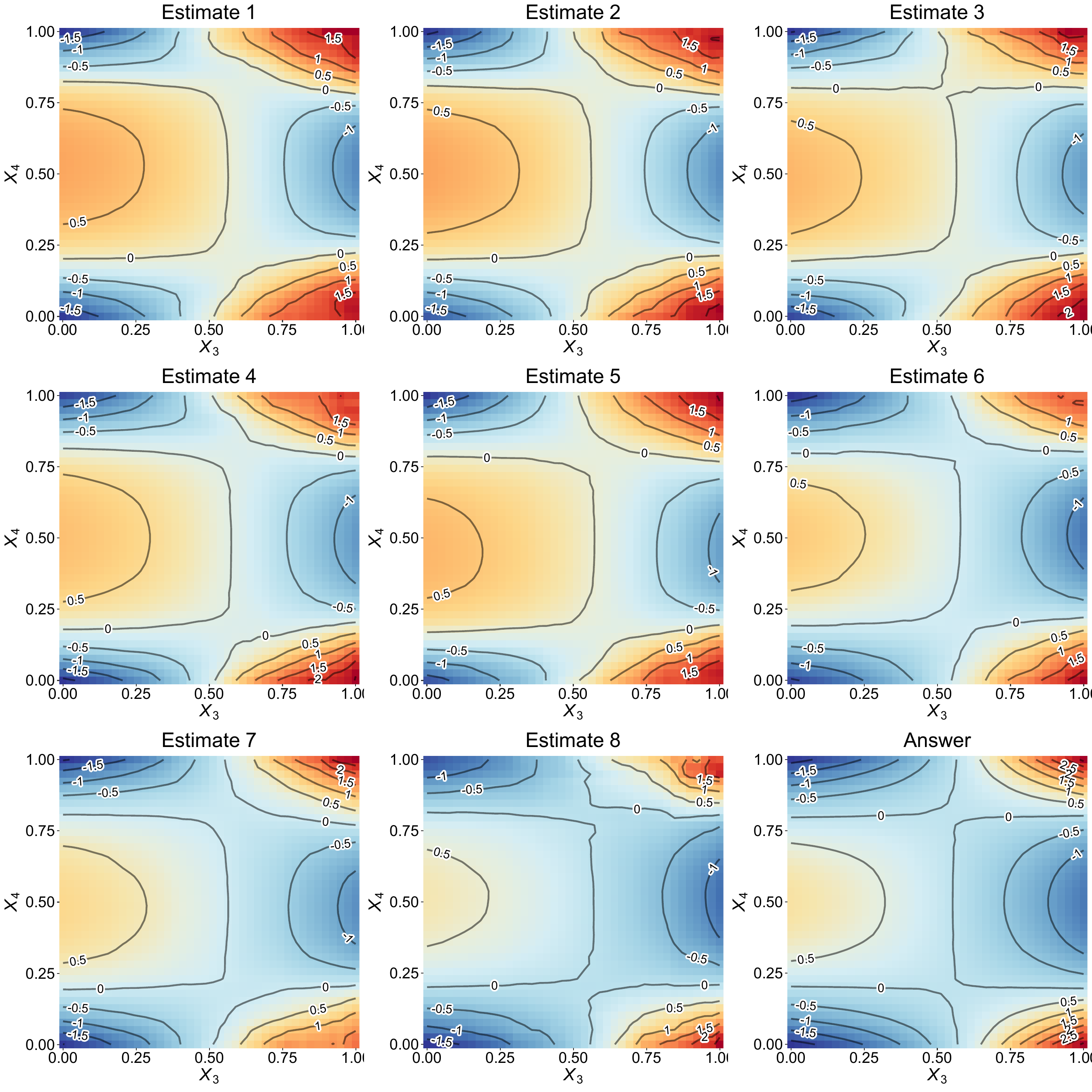}}
\caption{Marginal interaction effect between $X_3$ and $X_4$. Estimate 1--8 are posterior mean ALE interaction effect $\hat{\bar{Q}}^I_{34}(\tau,x_3,x_4)$ of 8 replicates at quantile level $\tau=0.5$. ``Answer" represents the ground truth.}\label{f:sens_int34_2}
\end{figure}

\begin{figure}
\centerline{\includegraphics[page=3,width=\linewidth]{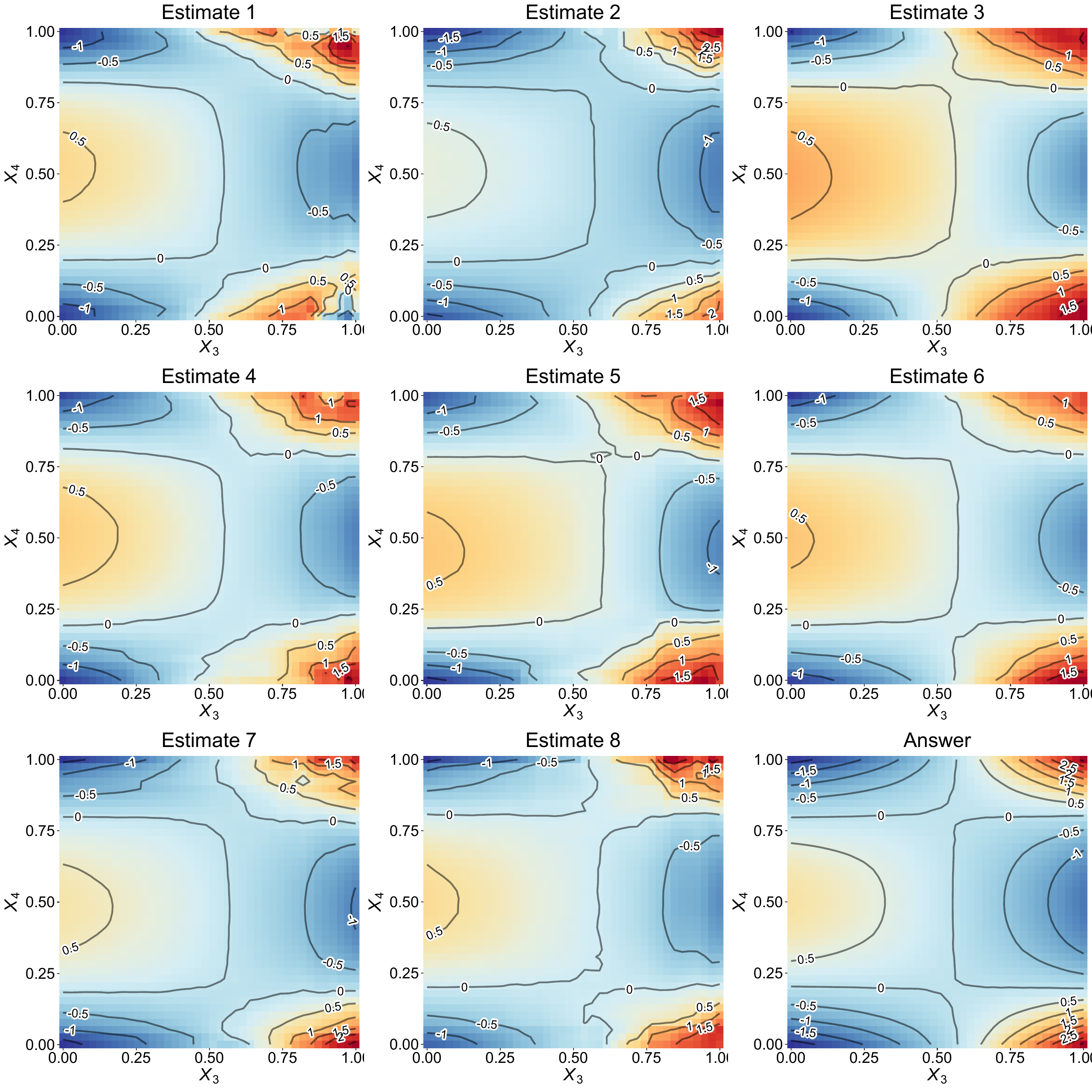}}
\caption{Marginal interaction effect between $X_3$ and $X_4$. Estimate 1--8 are posterior mean ALE interaction effect $\hat{\bar{Q}}^I_{34}(\tau,x_3,x_4)$ of 8 replicates at quantile level $\tau=0.95$. ``Answer" represents the ground truth.}\label{f:sens_int34_3}
\end{figure}

\begin{figure}
\centerline{\includegraphics[page=7,width=\linewidth]{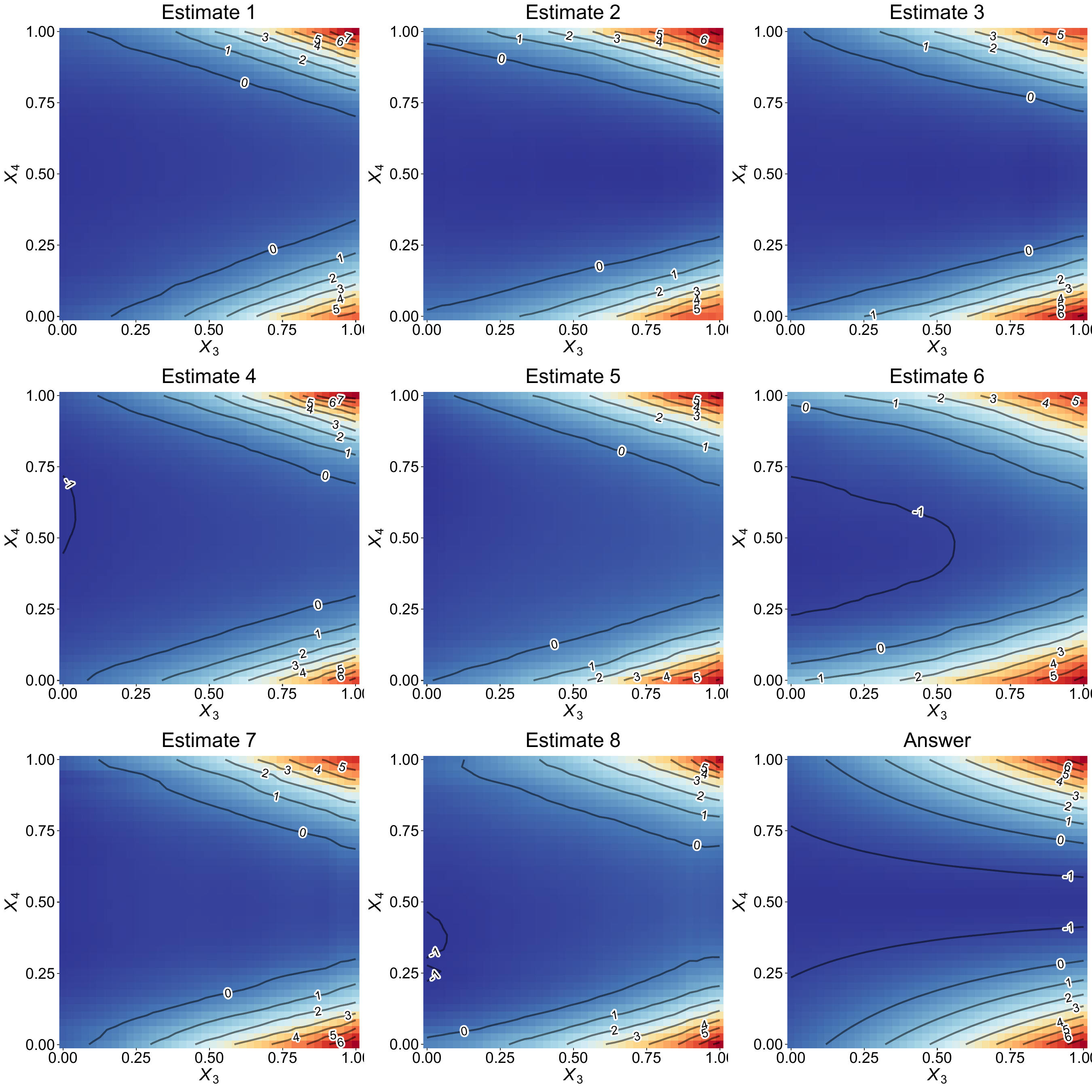}}
\caption{Marginal joint effect of $X_3$ and $X_4$. Estimate 1--8 are posterior mean ALE joint effect $\hat{\bar{Q}}_{34}(\tau,x_3,x_4)$ of 8 replicates at quantile level $\tau=0.05$. ``Answer" represents the ground truth.}\label{f:sens_int34_4}
\end{figure}

\begin{figure}
\centerline{\includegraphics[page=8,width=\linewidth]{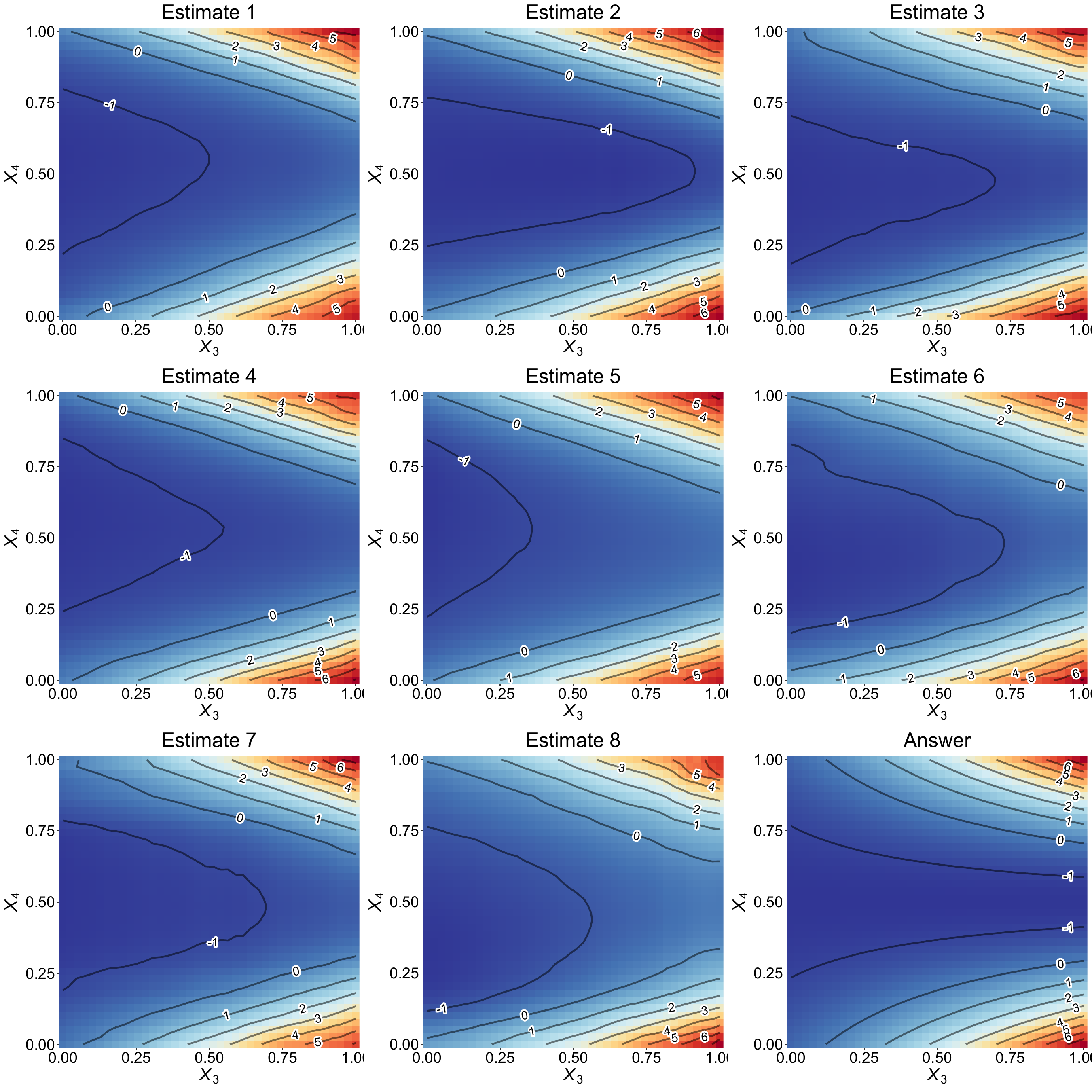}}
\caption{Marginal joint effect between $X_3$ and $X_4$. Estimate 1--8 are posterior mean ALE joint effect $\hat{\bar{Q}}_{34}(\tau,x_3,x_4)$ of 8 replicates at quantile level $\tau=0.5$. ``Answer" represents the ground truth.}\label{f:sens_int34_5}
\end{figure}

\begin{figure}
\centerline{\includegraphics[page=9,width=\linewidth]{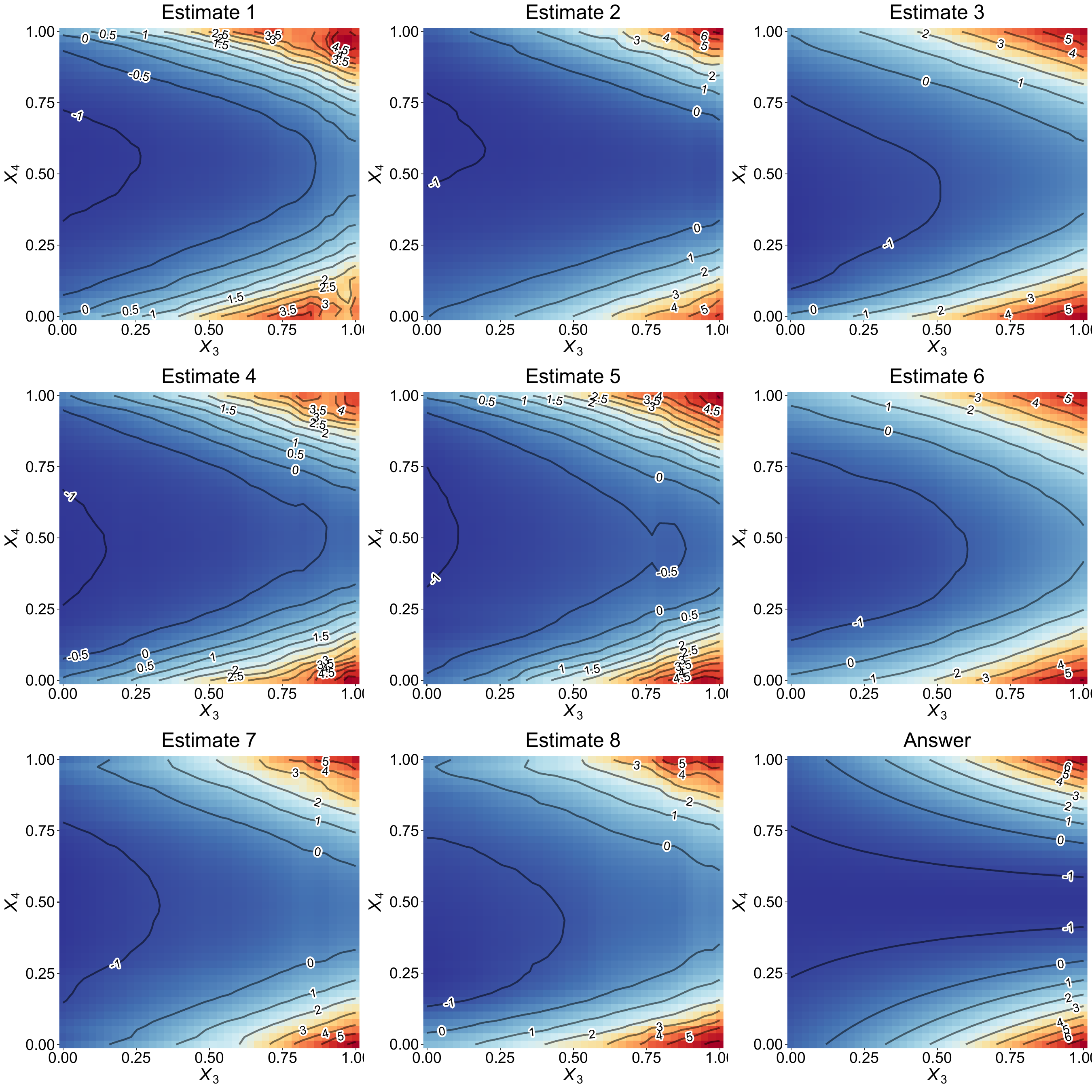}}
\caption{Marginal joint effect of $X_3$ and $X_4$. Estimate 1--8 are posterior mean ALE joint effect $\hat{\bar{Q}}_{34}(\tau,x_3,x_4)$ of 8 replicates at quantile level $\tau=0.95$. ``Answer" represents the ground truth.}\label{f:sens_int34_6}
\end{figure}

\begin{figure}
\centerline{\includegraphics[page=4,width=\linewidth]{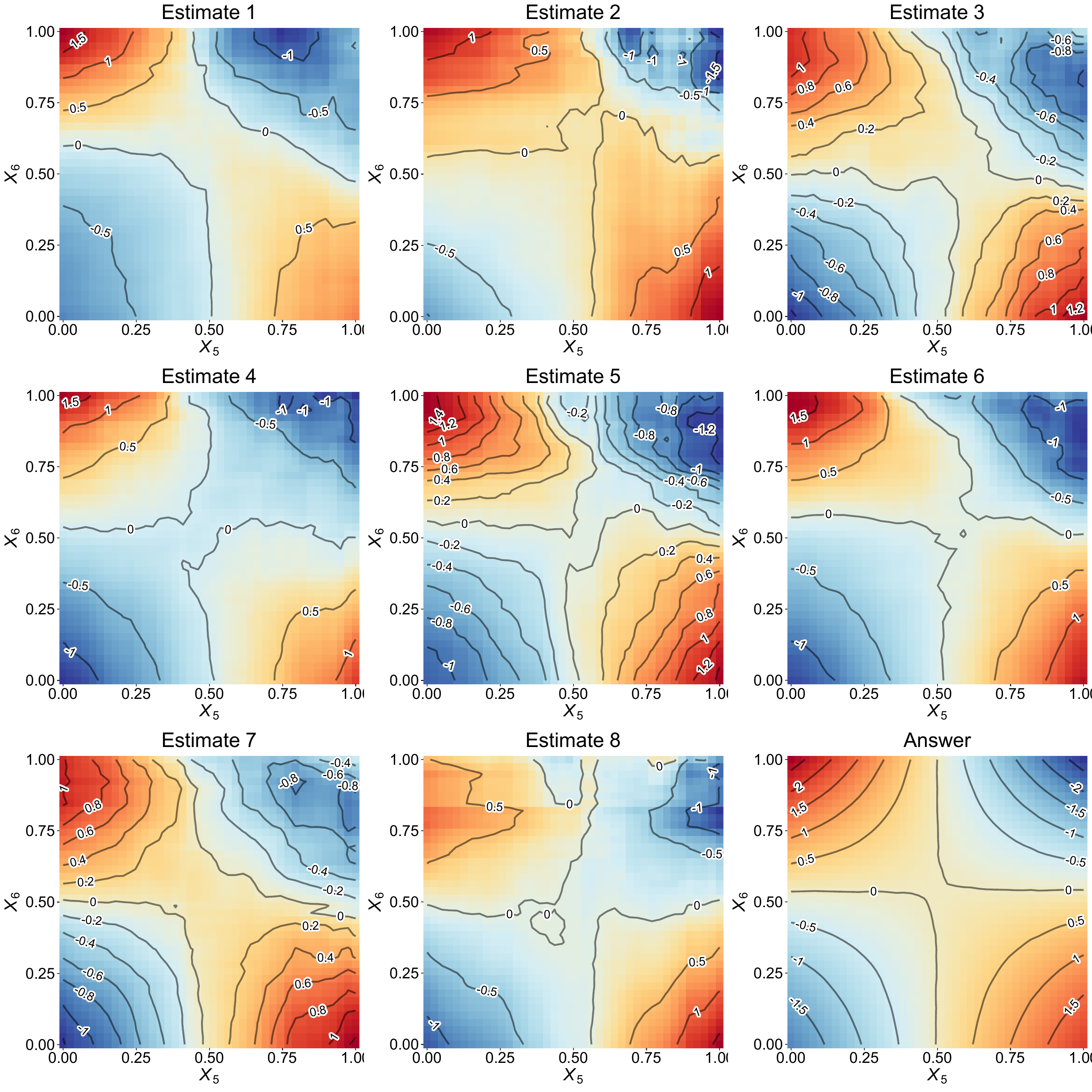}}
\caption{Marginal interaction effect between $X_5$ and $X_6$. Estimate 1--8 are posterior mean ALE joint effect $\hat{\bar{Q}}^I_{56}(\tau,x_5,x_6)$ of 8 replicates at quantile level $\tau=0.05$. ``Answer" represents the ground truth.}\label{f:sens_int56_1}
\end{figure}

\begin{figure}
\centerline{\includegraphics[page=5,width=\linewidth]{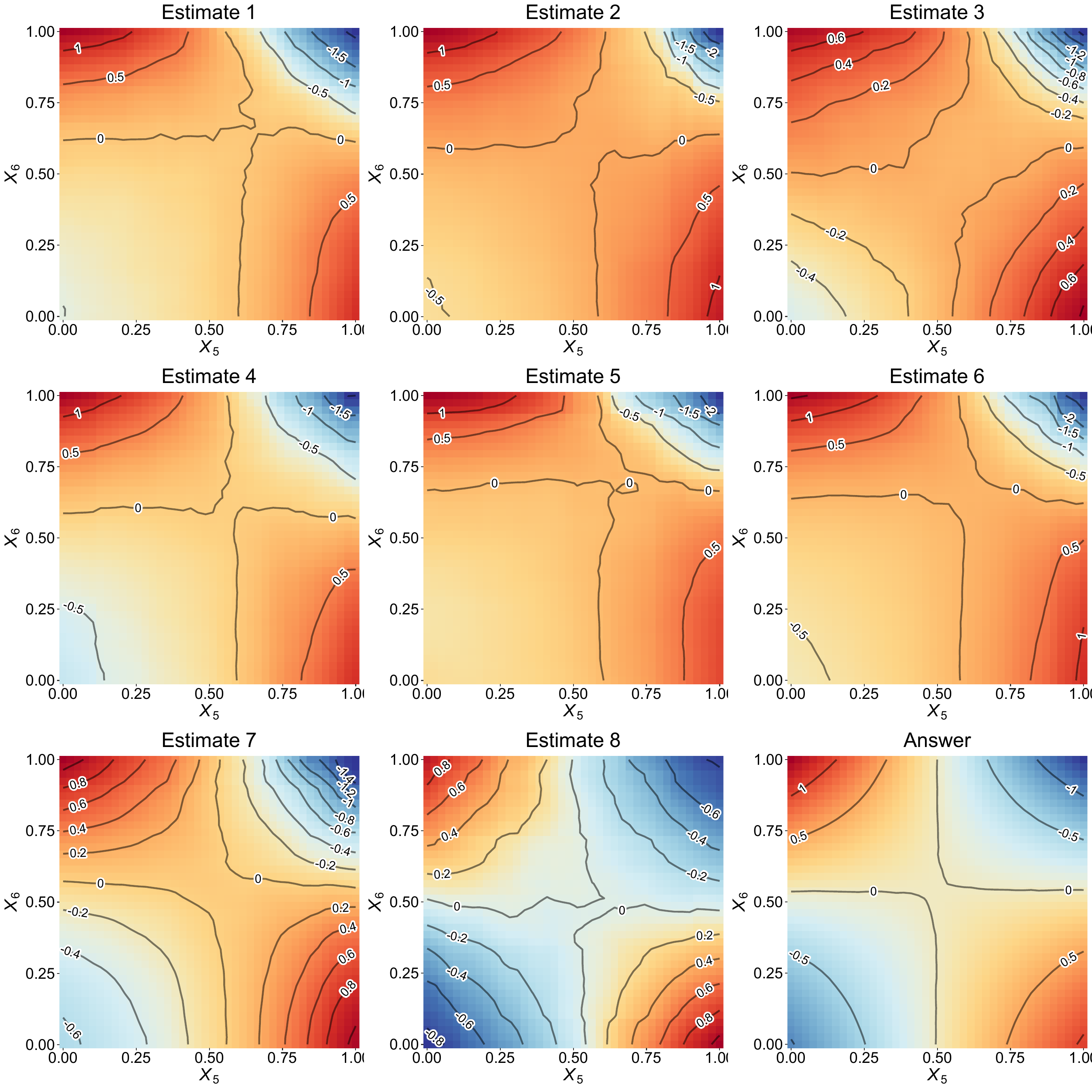}}
\caption{Marginal interaction joint of $X_5$ and $X_6$. Estimate 1--8 are posterior mean ALE interaction effect $\hat{\bar{Q}}^I_{56}(\tau,x_5,x_6)$ of 8 replicates at quantile level $\tau=0.5$. ``Answer" represents the ground truth.}\label{f:sens_int56_2}
\end{figure}

\begin{figure}
\centerline{\includegraphics[page=6,width=\linewidth]{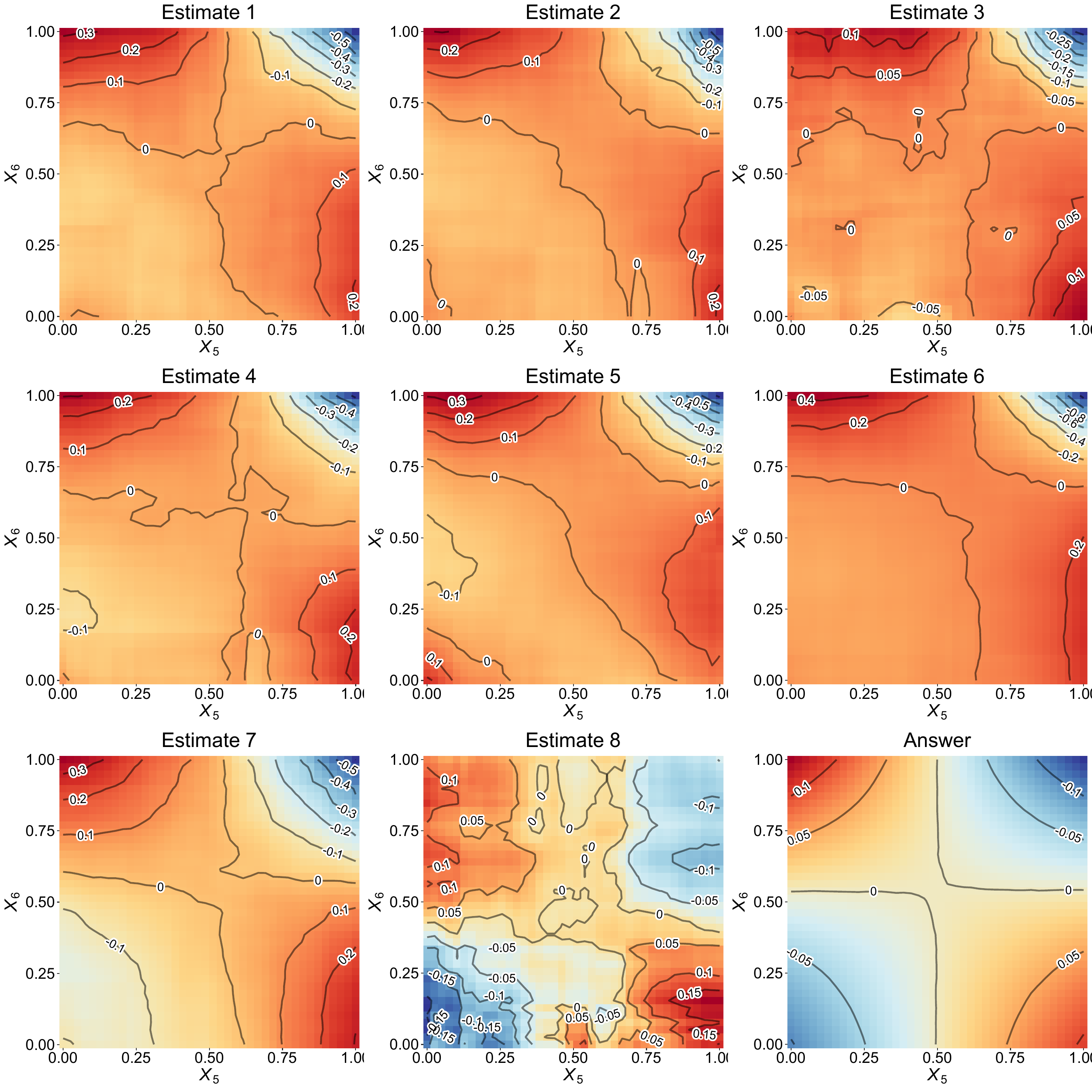}}
\caption{Marginal interaction effect between $X_5$ and $X_6$. Estimate 1--8 are posterior mean ALE interaction effect $\hat{\bar{Q}}^I_{56}(\tau,x_5,x_6)$ of 8 replicates at quantile level $\tau=0.95$. ``Answer" represents the ground truth.}\label{f:sens_int56_3}
\end{figure}

\begin{figure}
\centerline{\includegraphics[page=10,width=\linewidth]{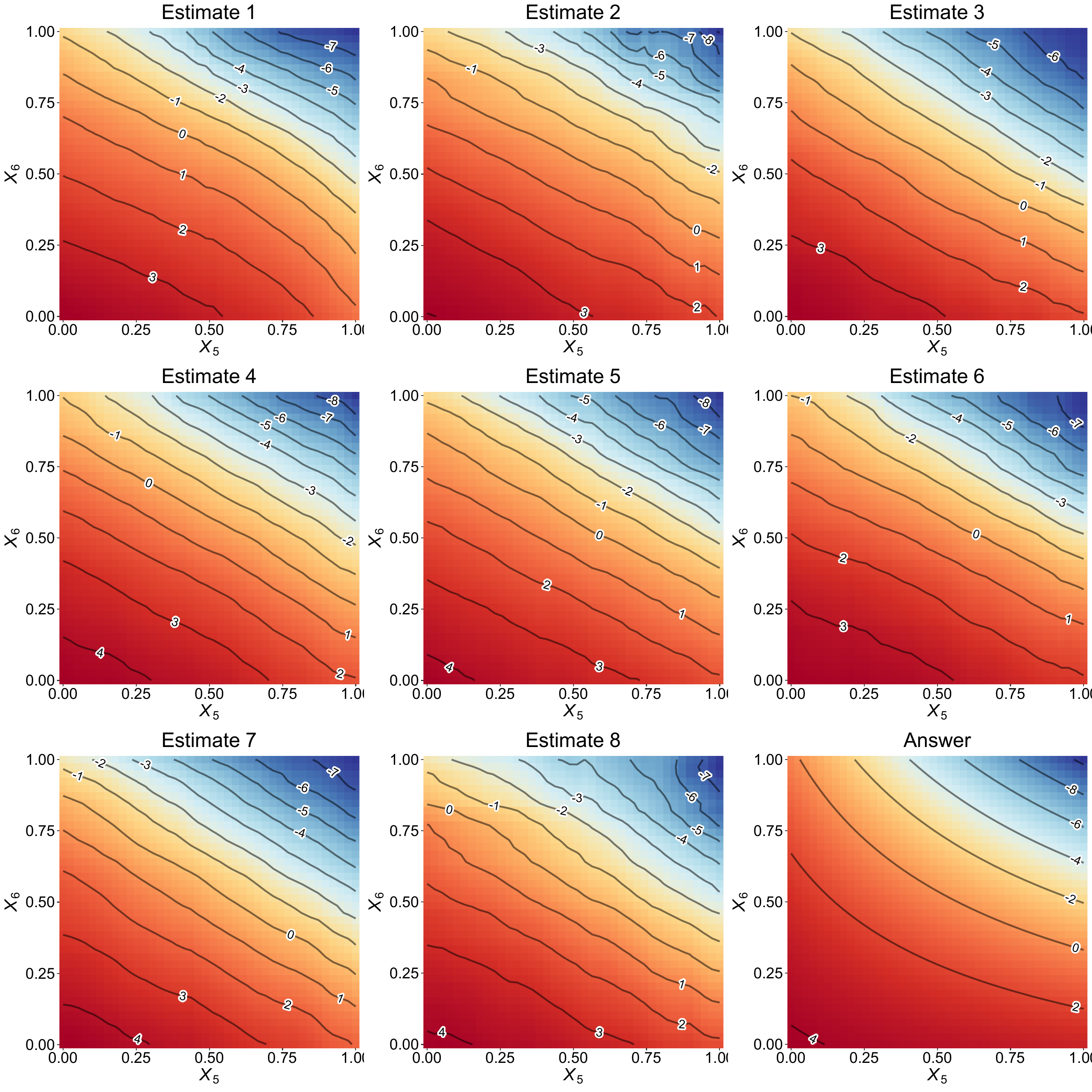}}
\caption{Marginal joint effect of $X_5$ and $X_6$. Estimate 1--8 are posterior mean ALE joint effect $\hat{\bar{Q}}_{56}(\tau,x_5,x_6)$ of 8 replicates at quantile level $\tau=0.05$. ``Answer" represents the ground truth.}\label{f:sens_int56_4}
\end{figure}

\begin{figure}
\centerline{\includegraphics[page=11,width=\linewidth]{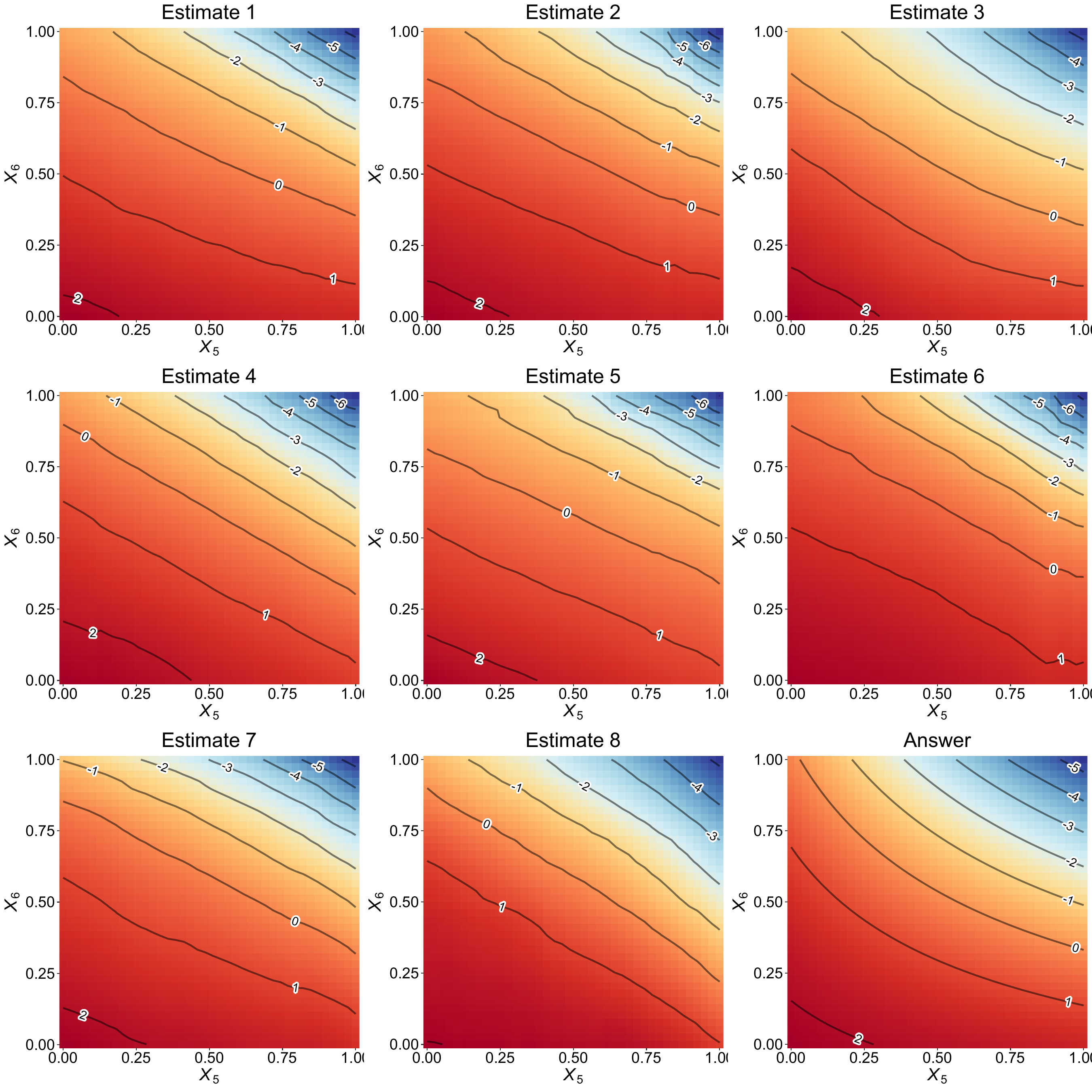}}
\caption{Marginal joint effect of $X_5$ and $X_6$. Estimate 1--8 are posterior mean ALE joint effect $\hat{\bar{Q}}^I_{56}(\tau,x_5,x_6)$ of 8 replicates at quantile level $\tau=0.5$. ``Answer" represents the ground truth.}\label{f:sens_int56_5}
\end{figure}

\begin{figure}
\centerline{\includegraphics[page=12,width=\linewidth]{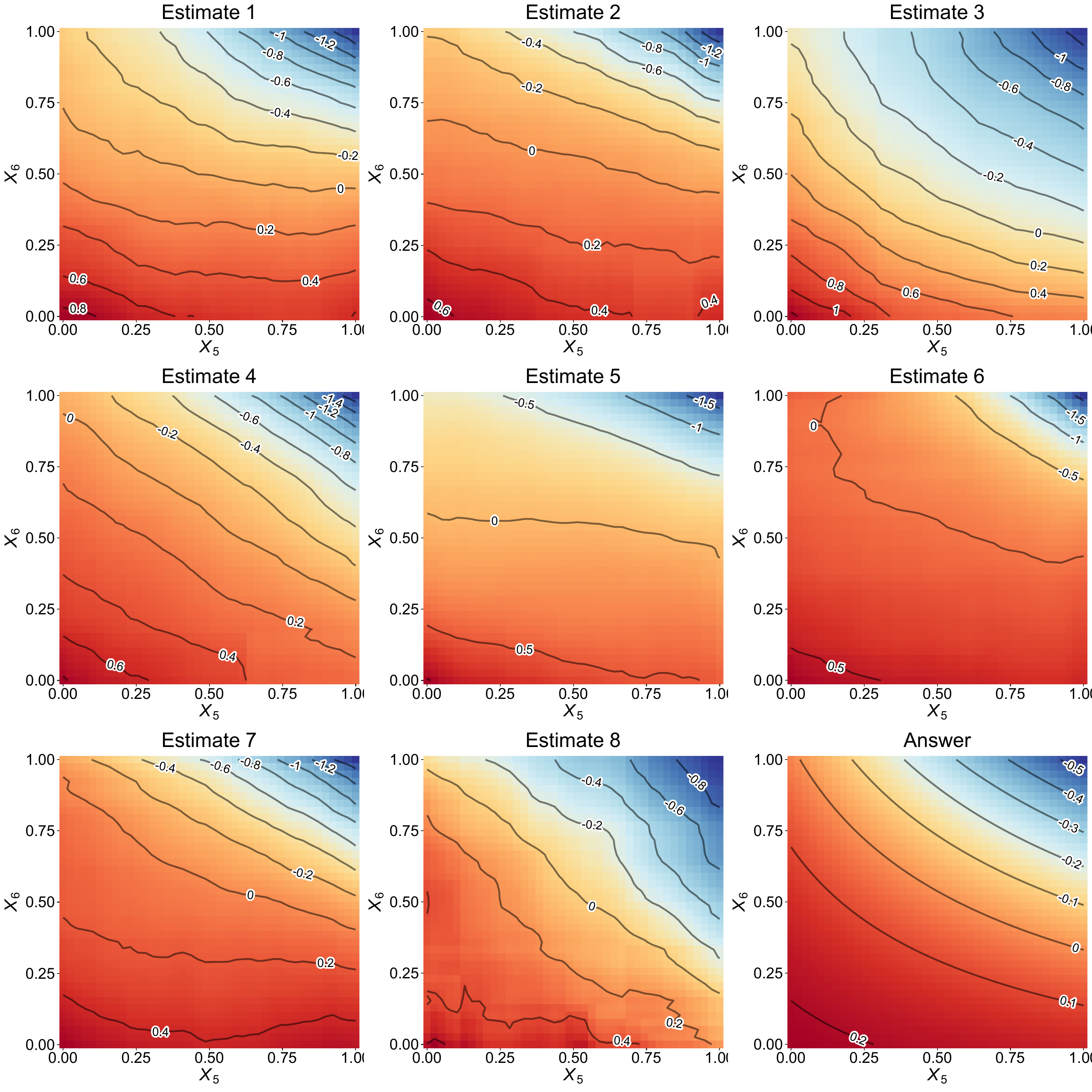}}
\caption{Marginal joint effect of $X_5$ and $X_6$. Estimate 1--8 are posterior mean ALE joint effect $\hat{\bar{Q}}^I_{56}(\tau,x_5,x_6)$ of 8 replicates at quantile level $\tau=0.95$. ``Answer" represents the ground truth.}\label{f:sens_int56_6}
\end{figure}

\end{document}